\documentclass[]{aa}

\usepackage[normalem]{ulem}
\usepackage{natbib}
\bibpunct{(}{)}{;}{a}{}{,}
\usepackage{scrextend}          
\usepackage{subcaption}
\usepackage{multirow}
\usepackage{amsmath}  
\usepackage{xfrac}   
\usepackage{amssymb}    
\usepackage{graphicx}
\usepackage{dsfont}
\usepackage{txfonts}
\usepackage{mwe,tikz}
\usepackage[percent]{overpic}
\usepackage{lipsum}
\usepackage{siunitx}
\usepackage{rotating}
\usepackage{hyperref}

\newcommand{\Rj}{$R_J$\xspace}
\newcommand{\Mdot}{$\dot{M}$\xspace}
\newcommand{\Mdotin}{$\dot{M}_{in}$\xspace}
\newcommand{\risco}{$r_{isco}$\xspace}
\newcommand{\mdotin}{$\dot{m}_{in}$\xspace}
\newcommand{\mdot}{$\dot{m}$\xspace}

\newcommand{\rj}{$r_J$\xspace}
\newcommand{\mdotint}{$\dot{m}_{in} (t)$\xspace}

\newcommand{\rjt}{$r_J (t)$\xspace}
\newcommand{\rjdemdotin}{$r_J \left(\dot{m}_{in} \right)$\xspace}
\newcommand{\rjdemdot}{$r_J \left(\dot{m} \right)$\xspace}

\newcommand{\pair}{$(r_J, \dot{m}_{in})$\xspace}

\newcommand{\ie}{i.e.\xspace}
\newcommand{\eg}{e.g.\xspace}

\newcommand{\jed}{jet-emitting disk\xspace}

\newcommand{\sad}{standard accretion disk\xspace}

\newcommand{\radeff}{radiative efficiency\xspace}
\newcommand{\radeffs}{radiative efficiencies\xspace}

\newcommand{\gx}{GX~339-4\xspace}

\newcommand{\one}{$\# 1$\xspace}
\newcommand{\two}{$\# 2$\xspace}
\newcommand{\three}{$\# 3$\xspace}
\newcommand{\four}{$\# 4$\xspace}

\begin{document} 

  \title{A unified accretion-ejection paradigm for black hole X-ray binaries}
  \subtitle{VI. Radiative efficiency and radio--X-ray correlation during four outbursts from \gx }

  \author{
            G. Marcel \inst{1} \fnmsep \inst{2}
            \and
                J. Ferreira \inst{3}
            \and
            P-O. Petrucci\inst{3}
            \and
            S. Barnier\inst{3}
            \and
            J. Malzac\inst{4}
            \and
            A. Marino\inst{5} \fnmsep \inst{6}
            \and
            M. Coriat\inst{4}
            \and
            M. Clavel\inst{3}
            \and
            C. Reynolds\inst{1}
            \and
            J. Neilsen\inst{2}
            \and
            R. Belmont\inst{7}
            \and
            S. Corbel\inst{7} \fnmsep \inst{8}
        }

   \institute{Institute of Astronomy, University of Cambridge, Madingley Road, Cambridge, CB3 OHA, United Kingdom\\
                \email{greg.marcel@cam.ac.uk or gregoiremarcel26@gmail.com}
                \and
                Villanova University, Department of Physics, Villanova, PA 19085, USA
                \and
                Univ. Grenoble Alpes, CNRS, IPAG, 38000 Grenoble, France
                \and
                IRAP, Université de Toulouse, CNRS, UPS, CNES, Toulouse, France
                \and 
                Universitá degli Studi di Palermo, Dipartimento di Fisica e Chimica, via Archirafi 36, I-90123 Palermo, Italy.
                \and
                INAF/IASF Palermo, via Ugo La Malfa 153, I-90146 - Palermo, Italy.
                \and
                AIM, CEA, CNRS, Université Paris-Saclay, Université Paris Diderot, Sorbonne Paris Cité, 91191 Gif-sur-Yvette, France
                \and
                Station de Radioastronomie de Nançay, Observatoire de Paris, PSL Research University, CNRS, Univ. Orléans, 18330 Nançay, France
              }

   \date{Received May 26, 2021; accepted September 27, 2021}

 \abstract{The spectral evolution of transient X-ray binaries (XrBs) can be reproduced by an interplay between two flows separated at a transition radius \Rj : a \sad (SAD) in the outer parts beyond \Rj and a \jed (JED) in the inner parts. In the previous papers in this series we successfully recover the spectral evolution in both X-rays and radio for four outbursts of \gx by playing independently with the two parameters: \Rj and the disk accretion rate \Mdotin . In this paper we compare the temporal evolution of both \Rj and \Mdotin for the four outbursts. We show that despite the undeniable differences between the time evolution of each outburst, a unique pattern in the $\dot{M}_{in}-R_J$ plane seems to be followed by all cycles within the JED-SAD model. We call this pattern a fingerprint, and show that even the ``failed'' outburst considered follows it. We also compute the radiative efficiency in X-rays during the cycles and consider its impact on the radio–X-ray correlation. Within the JED-SAD paradigm, we find that the accretion flow is always radiatively efficient in the hard states, with between $15\%$ and $40\%$ of the accretion power being radiated away at any given time. Moreover, we show that the radiative efficiency evolves with the accretion rate because of key changes in the JED thermal structure. These changes give birth to two different regimes with different radiative efficiencies: the thick disk and the slim disk. While the existence of these two regimes is intrinsically linked to the JED-SAD model, we show direct observational evidence of the presence of two different regimes using the evolution of the X-ray power-law spectral index, a model-independent estimate. We then argue that these two regimes could be the origin of the gap in X-ray luminosity in the hard state, the wiggles, and different slopes seen in the radio--X-ray correlation, and even the existence of outliers.}
 
   \maketitle
%

\section{Introduction}

\begin{figure*}[h!]
\begin{center}
  \includegraphics[width=1.0\linewidth]{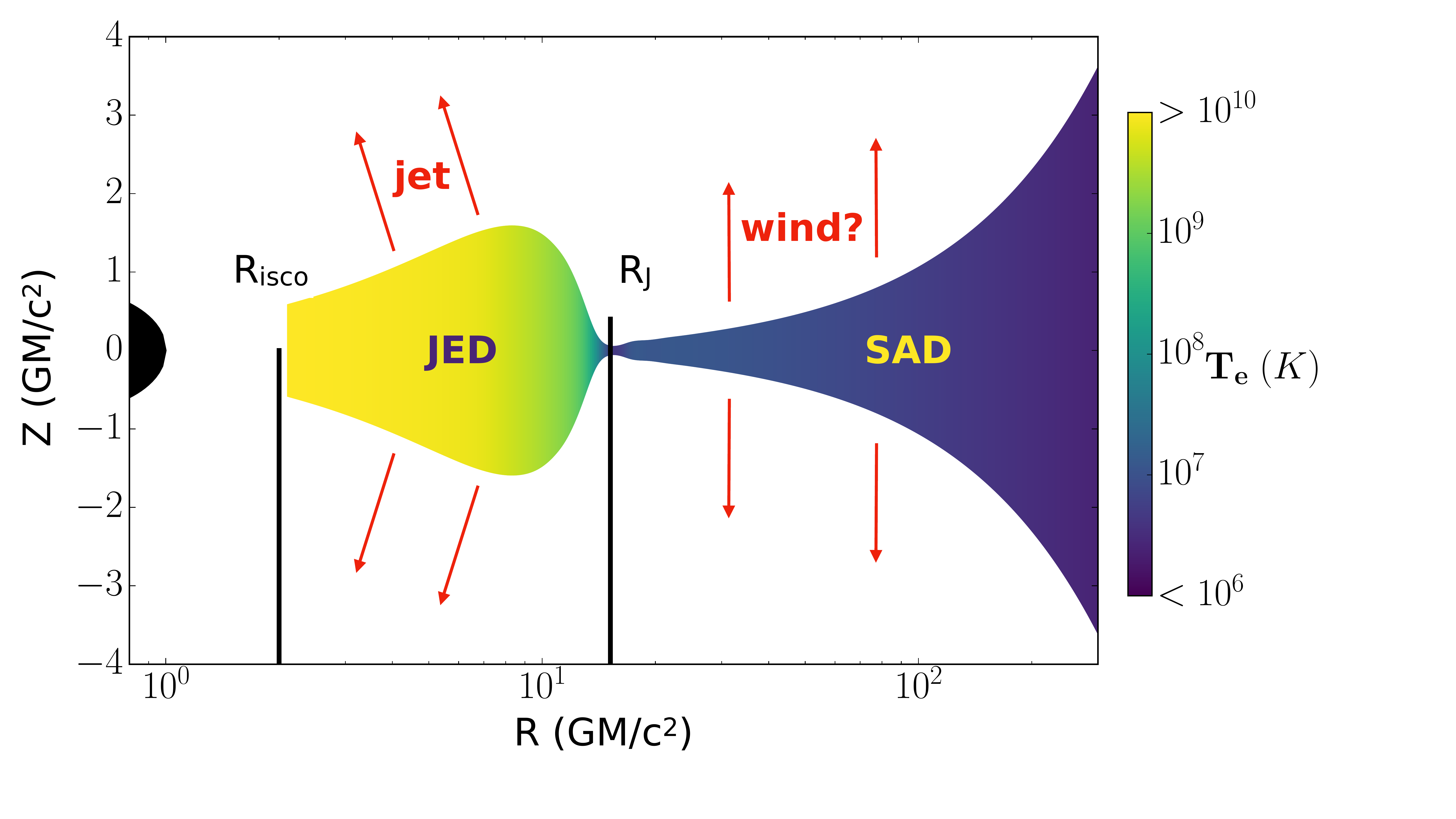}
   \caption{Example of solution showing the geometry of the JED-SAD paradigm. From left to right: Black hole (in black), \jed from $R_{isco}$ to \Rj (in yellow--green), and \sad beyond \Rj (violet). This example is an actual physical calculation, where the color is the actual electron temperature, adapted from Fig.~1 in \citet{Marcel18b}.}
  \label{fig:config}
  \end{center}
\end{figure*}

Black hole X-ray binaries are composed of a stellar mass black hole and a companion star. Over time, matter from the companion accretes onto the black hole to form an accretion flow or accretion disk \citep[for a review, see][]{Remillard06}. While X-ray binaries spend most of their life in a quiescent and barely detectable state, they often undergo huge outbursts usually detected in X-ray \citep{Dunn10}. All sources have different behaviors; some undergo outbursts every other year, while others have been stuck in outburst since their discovery \citep[see][]{Tetarenko16}. Each outburst from a given source is unique, but most follow a similar evolution through two distinct spectral states observed in X-rays: hard and soft. The hard state is characterized by a Comptonization spectrum mainly in the hard X-rays (above $10$\,keV), while the soft state is characterized by a disk blackbody with typical temperature in the soft X-rays (below $1$\,keV). During a given outburst, a source will start in the quiescent state and rise in luminosity in the hard state for up to $3-4$ orders of magnitude \citep[see, however, failed outbursts;][and references therein]{Tetarenko16}. Once it reaches Eddington-like luminosities in X-ray $\gtrsim 10\% \, L_{Edd}$ the source transitions to the soft state, where luminosity will gradually decrease. When the luminosity reaches about $1-5\% \, L_{Edd}$, a luminosity that seems constant for all outbursts from a given source \citep{Maccarone03}, the source transitions back to the hard state where it will eventually return to the quiescent state. In this common behavior, transition phases between hard and soft usually last a few days, while both the hard and the soft phases can last months. In addition to these changes in X-ray, we observe drastic variations at other wavelengths, especially in radio bands. During the hard state, a weak radio counterpart is usually detected. However, it disappears (is quenched) entirely when the system reaches the soft state. A key correlation has been discovered in the hard state between the radio (around $5-9$\,GHz) and the soft X-rays \citep[usually $1-10$\,keV or $3-9$\,keV;][]{1998NewAR..42..601H, Corbel03, Gallo03}. There are two major tracks in this correlation, labeled standard and outliers for historical reasons, and their origin is still unknown \citep{2012MNRAS.423..590G, Corbel13, 2014MNRAS.440..965H, 2018MNRAS.478L.132G}, although possible differences in the jets \citep{2018MNRAS.473.4122E} or the accretion flow properties have been mentioned \citep{Coriat11, Koljonen19}. In addition to these two jets, strong winds are usually detected, especially in the soft states \citep[see, e.g.,][]{Ponti12}. However, it is still unclear exactly when these winds are produced and observed \citep[see introduction in][]{2021A&A...649A.128P}.

The behavior depicted above is quite generic and well characterized, but there is still no consensus explanation about what causes a cycle and what drives the evolution in X-rays and radio \citep{Done07, Yuan14}. A unified framework, the jet-emitting disk--standard accretion disk (JED-SAD) paradigm has been progressively developed to address these points in a series of papers. The framework was proposed by \citeauthor{Ferreira06} (\citeyear{Ferreira06}, hereafter paper~I). They assume that the accretion disk is threaded by a large-scale magnetic field, whose radial distribution separates the disk into two different accretion flows. Outside a radius \Rj , in the outer region, the disk is barely magnetized, in the regime that they call a standard accretion disk \citep[SAD,][]{SS73}. In this regime, accretion is mainly due to turbulence, through what we now think is the magneto-rotational instability \citep[MRI,][]{BH91,Balbus03}. Although ejections (winds or jets) can be produced in these conditions, they have been neglected in this paradigm so far, as we assume that only the inner region produces ejections (see Fig.~\ref{fig:config} and below). In its inner region, from the inner-most stable circular orbit (ISCO) $R_{isco}$ to the transition radius \Rj , the disk is magnetized around equipartition (\ie, the magnetic pressure is approximately the sum of the radiative and gaseous pressures). The presence of a strong vertical magnetic field allows for the production of powerful and self-confined ejections \citep{BP82}, and these ejections apply a torque on the disk that accelerates matter. This regime is called a jet-emitting disk \citep[JED,][]{Ferreira93a, Ferreira93b, Ferreira95}. We show an example of JED-SAD configuration in Fig.~\ref{fig:config}, adapted from \citet{Marcel18b}. We invite the interested reader to study the previous papers in this series (see next paragraph), the seminal papers of the JEDs \citep{Ferreira93a, Ferreira93b, Ferreira95}, as well as the latest numerical calculations \citep{2019MNRAS.490.3112J, 2021A&A...647A.192J} and related simulations \citep{Scepi19, Liska19}. 

In order to compare this framework to observations, \citeauthor{Marcel18a} (\citeyear{Marcel18a}, hereafter paper~II) designed a two-temperature plasma code that computes the thermal structure and its associated spectral emission for a JED (\ie, \Rj is pushed to infinity in Fig.~\ref{fig:config}). They showed that a JED could be the source of the hard X-rays (\ie, the region commonly called corona) in a given set of parameters. These parameters are either linked to the micro-physics or global parameters such as the black hole spin or mass, or even the distance to the source. The parameter set was chosen to be physically consistent with numerical calculations \citep{2019MNRAS.490.3112J,2021A&A...647A.192J}, observations \citep{Petrucci10}, and the known parameters of the source \gx \citep{Marcel18a}. They then added an outer standard accretion disk (SAD) outside of the radius \Rj \citep[see Fig.~\ref{fig:config} and][hereafter paper~III]{Marcel18b}, essentially making \Rj a parameter in the model. They froze all the parameters in the model to their expected value from paper~II, and showed that the canonical states observed in X-ray binaries could be reproduced by playing independently with only two remaining parameters: the transition radius \Rj and the accretion rate in the disk \Mdotin . It is important to note here that the model does not need nor does it include any normalization. The X-ray flux is directly dependant on the thermal state of the disk (e.g., temperature, density, optical depth) as well as the distance to the source. Later, they qualitatively reproduced four full cycles of \gx , playing again only with \Rj and \Mdot (\citeauthor{Marcel19} \citeyear{Marcel19}, paper~IV; \citeauthor{2020A&A...640A..18M} \citeyear{2020A&A...640A..18M}, paper~V). Their spectral fitting method (see Sect.~\ref{sect:meth}) includes both radio and X-rays, and compared the results to the presence of quasi-periodic oscillations. We invite the reader to read the previous papers in the series for more detailed discussions about the theoretical framework (paper~I), other models (paper~II, Introduction), the assumptions and equations resolved (papers~II, III), the fitting procedure and the inclusion of radio flux (paper~IV), and the comparison to timing properties (paper~V).

The goal of the present study is to build up a generic picture of the archetypal source \gx within the JED-SAD paradigm. We then use this generic picture and tackle questions about the radiative efficiency of the accretion flow and the radio--X-ray correlation. In section~\ref{sec:fingerprint} we describe the methodology used, and show that all four outbursts from \gx follow a similar path in the theoretical $\dot{M}_{in}-R_J$ plane despite clear differences between outbursts. In section~\ref{sec:diskstruc} we focus on the evolution of the disk structure within the JED-SAD paradigm, and we address the existence of two different accretion regimes during the hard state. We then discuss the evolution of the \radeff of the accretion flow in Sect.~\ref{sec:secradeff}, as well as possible observational evidence for these different regimes. We also discuss the impact on the radio--X-ray correlation. We detail the important possible caveats in section~\ref{sec:caveatsradio}, before discussing and concluding in section~\ref{sec:Ccl}.

\begin{figure*}[t!]
\begin{center}
  \includegraphics[width=1.0\linewidth]{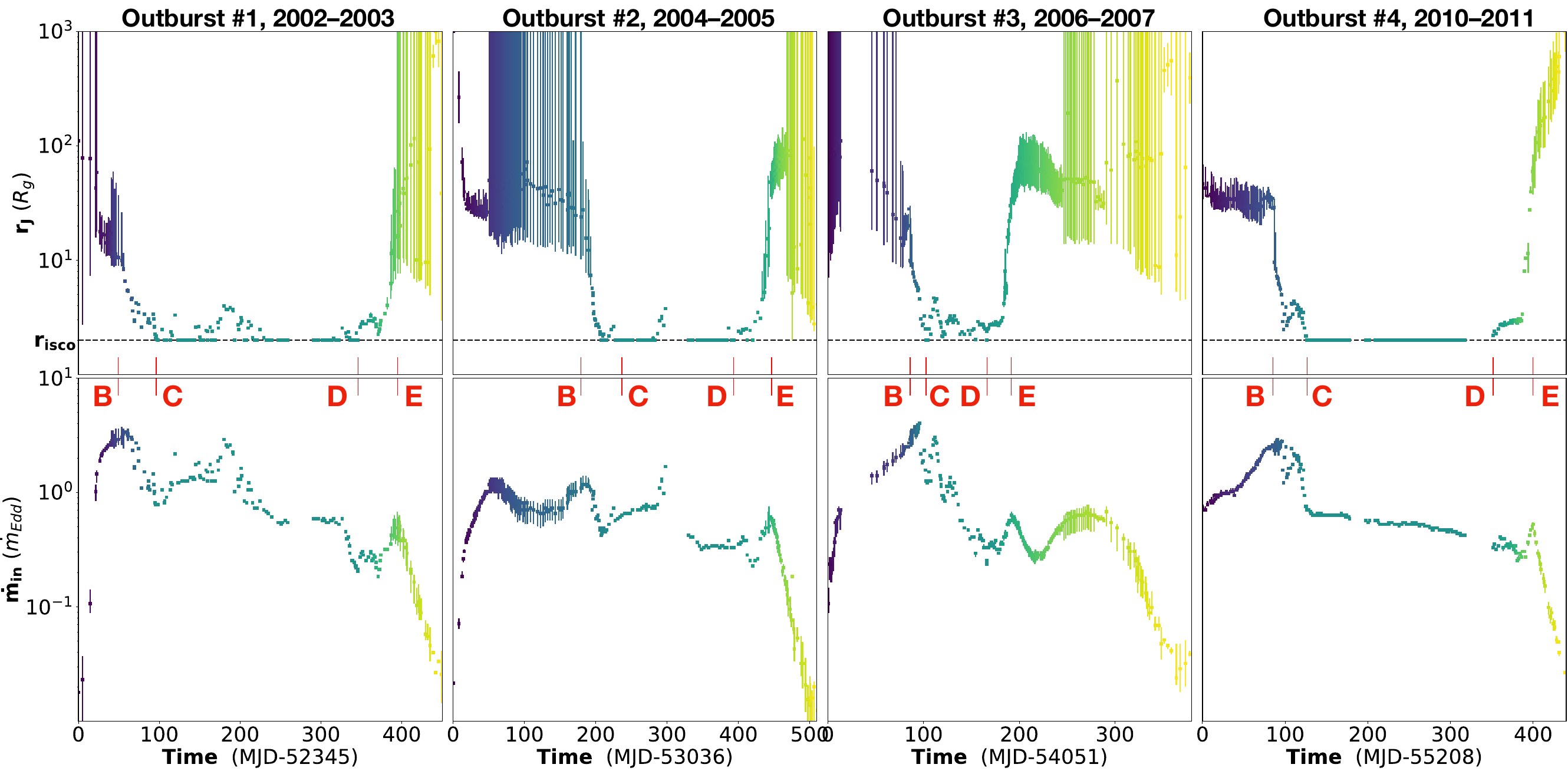}
   \caption{Evolution along time of \rj (top) and \mdotin (bottom) for all four outbursts of \gx observed by \text{RXTE} with their $5\,\%$ confidence regions. For each outburst the timescale is chosen so that $t=0$ corresponds to the first detection of the outburst (see paper~V for more details). The color-coding translates the time evolution for each outburst: starting in dark violet during the rising hard state and finishing in light yellow in the decaying hard state, using a constant color during the entire soft and soft-intermediate states. Four letters are placed to guide the evolution of each outburst: Point~B is the last hard state of the rising phase, C and D are the first and last soft states, and E is the first hard state of the decaying phase.}
  \label{fig:RjMdot1234}
  \end{center}
\end{figure*}

\section{A unique fingerprint for \gx } \label{sec:fingerprint}

\subsection{Methodology} \label{sect:meth}

The hybrid disk configuration is composed of a black hole of mass $M$, an inner jet-emitting disk (JED) from the inner stable circular orbit $R_{isco}$ to the transition radius $R_J$, and an outer standard accretion disk (SAD) from $R_J$ to $R_{out}$. The system is assumed to be at a distance $D$ from the observer. In the following we adopt the dimensionless scalings $r = R/R_g$, where $R_g = GM/c^2$ is the gravitational radius; $m = M/M_\odot$; and the local disk accretion rate $\dot{m} = \dot{M} /\dot{M}_{Edd}$, where $\dot{M}_{Edd}= L_{Edd}/c^2$ is the Eddington accretion rate and $L_{Edd}$ is the Eddington luminosity \citep{Eddington}. Because jets carry matter away from the disk, the accretion rate in a JED varies with radius $\dot{m} \propto r^{\xi}$ \citep{Ferreira95, BB99}, where the ejection efficiency $\xi$ is sometimes labeled $p$ or $s$ to avoid confusion with the ionization parameter. We use $\xi = 0.01$ in our analysis, consistent with the work from previous papers in this series and the most recent self-similar calculations on the issue \citep[see, e.g.,][Figure 7]{2019MNRAS.490.3112J}. We also mainly refer to the accretion rate at the ISCO, \mdotin , leading to $\dot{m} (r) = \dot{m}_{in} \, (r/r_{isco})^{\xi}$ at any given radius $r \leq r_J$ in the JED, and $\dot{m} (r) = \dot{m}_{in} \, (r_J/r_{isco})^{\xi}$ at any given radius $r \geq r_J$ in the SAD. In practice the accretion rate is almost constant since we use $\xi \ll 1$. Since we focus on the archetypal object \gx, we use a black hole mass $m= 5.8$, a spin $a = 0.93$ corresponding to $r_{isco} = 2.0$, and a distance $D = 8$ kpc \citep[][for more recent estimates]{Miller04, Munoz08, Parker16, Heida17}. The impact of the different parameters (e.g., $\xi$, $m_s$, $r_{isco}$) is thoroughly discussed in the previous papers in the series.

Four outbursts of \gx , from 2002 to 2011, were successfully recovered by playing with only the two (assumed) independent parameters $r_J$ and $\dot{m}_{in}$.
Each observation consists in a $3-25$\,keV X-ray spectral energy distribution (\textit{RXTE}/PCA) that we compare to our global theoretical spectral energy distribution. Some observations are also accompanied by a radio flux density at $8.6$ or $9$\,GHz (ATCA) that we compare to our radio flux estimate using
\begin{equation}
    F_R = \tilde{f}_R \dot{m}_{in}^{17/12} r_{isco} \left( r_J - r_{isco} \right)^{5/6} F_{Edd} \label{eq:radioflux}
\end{equation}
with $\tilde{f}_R$ the normalization factor (paper~IV). For each observation, we find the parameter pair $(r_J, \dot{m}_{in})(t)$ that best reproduces the main X-ray spectral parameters and the radio flux (when observed within one day of the X-ray). We refer the reader to the previous papers in this series for further details on the spectral parameters chosen, the fitting method, and the radio estimates (section~3 in paper~IV), as well as the spectral coverage in radio and X-rays (Figs.~1, 2, and 3 in paper~V).

\subsection{Temporal evolution} \label{sec:timevol}

We show in Fig.~\ref{fig:RjMdot1234} the evolution of \rjt and \mdotint, along with their $5 \%$ confidence regions for the four different outbursts from left to right: 2002-2003 (\one), 2004-2005 (\two), 2006-2007 (\three), and 2010-2011 (\four). These confidence regions describe the intervals for $r_J$ and $\dot{m}_{in}$, where any variation will lead to a change of at most $5 \%$ in the best fit. In paper~IV, especially section 3.2.1, we discuss these error bars and their importance. We use a unique color-scale for each outburst to show its unique time-evolution: an outburst starts in dark violet during the rising hard state and finishes in light yellow in the decaying hard state. We use a classical definition of the spectral states: quiescent, hard, hard- and soft-intermediate, and soft states (see paper~IV for their precise definitions). For ease of comparison between the various outbursts, we define five labels: A, B, C, D, and E, similarly to what was done by \citet{Petrucci08} and \citet{Kylafis15}. We define Point~A as the quiescent state, not visible in Fig.~\ref{fig:RjMdot1234}. We define Point~B as the last hard state of the rising phase, C and D as the first and last soft states, and E as the first hard state of the decaying phase. While the position of each label depends on the arbitrary definition chosen for each state (\ie, hard, hard-intermediate, soft-intermediate, or soft) they are convenient to help follow a given outburst.

In our paradigm, each outburst undergoes the following steps: a rise in \mdotin in the hard-state until B, a decrease in \rj in the transition to the soft-state where \rj reaches the ISCO until C, a gradual decrease in \mdotin in the soft-state where \rj stays at the ISCO until D, an increase in \rj to come back to the hard-state until E, and a decrease in \mdotin to go back to quiescence. In Fig.~\ref{fig:RjMdot1234} the two parameters \pair seem to vary independently: \rj undergoes its swiftest variations while \mdotin remains constant, and vise versa. This typical behavior is especially visible in outburst \four , where variations are much clearer and smoother thanks to smaller confidence regions due to the presence of radio flux in the fits. This is the behavior expected in the qualitative picture proposed in paper~I, but also in \citealt{Esin97}, although these authors make different assumptions and do not take into account the effect of jets on the accretion flow dynamics. We note that there are a few alternative scenarios trying to explain state transitions using either evaporation processes \citep{Meyer94, Meyer05}, a disk dynamo \citep{BegelmanArmitage14}, or the cosmic battery \citep{Kylafis15}. 

\begin{figure*}[t!]
\begin{center}
  \includegraphics[width=0.93\linewidth]{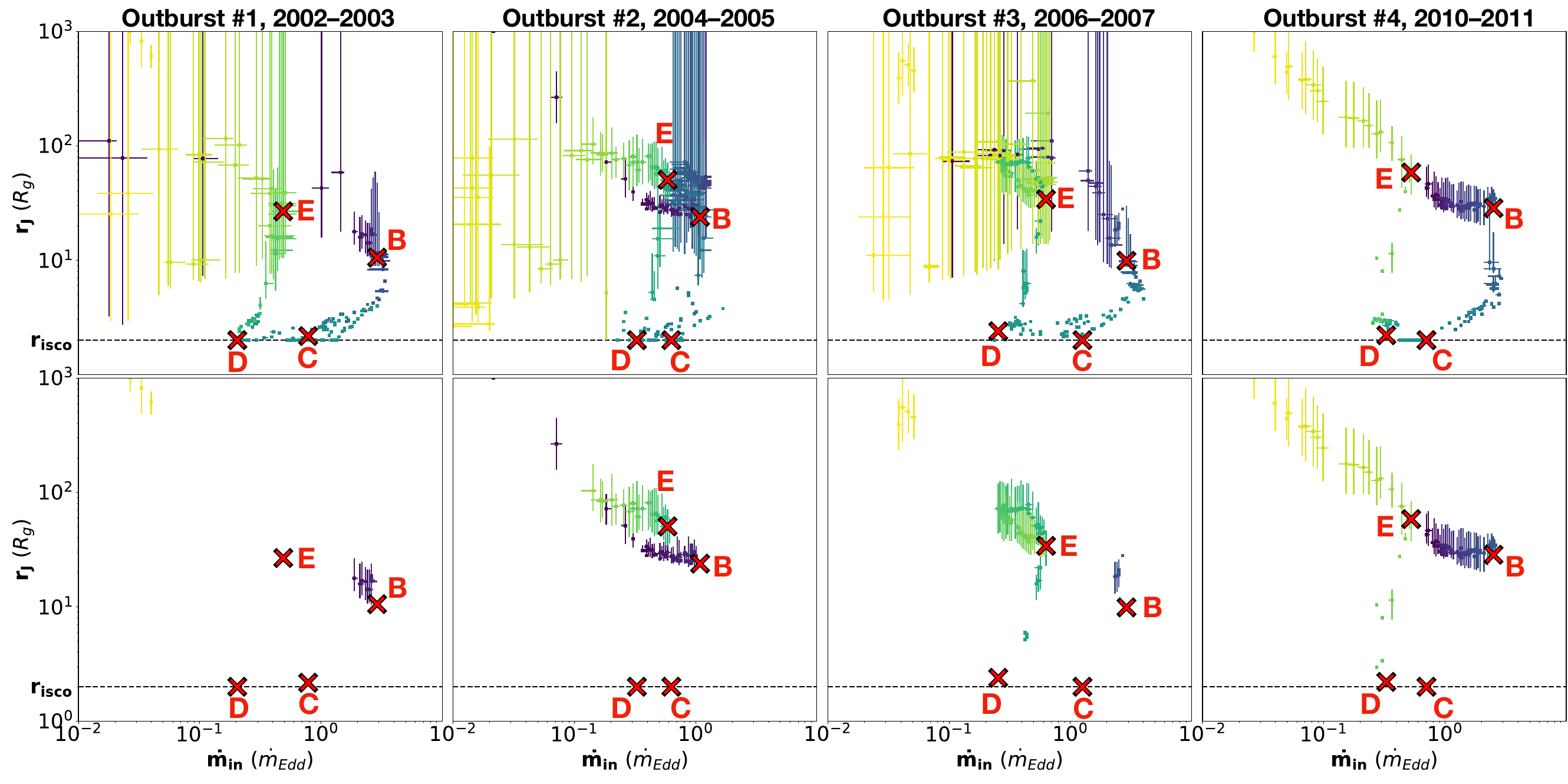}
   \caption{Positions of each observation in the $\dot{m}_{in}-r_J$ plane for each outburst: all observations (top), only those when radio constraints are present (bottom). The colors are the same as in Fig.~\ref{fig:RjMdot1234}: each outburst starts in the dark violet phase, transitions to the soft state in green, and comes back to the hard state in the light yellow phase. Additionally, the labels B, C, D, and E from Fig.~\ref{fig:RjMdot1234} are shown.}
  \label{fig:RjMdot1234_bis}
  \end{center}
\end{figure*}

While all outbursts follow the same generic evolution (A-B-C-D-E-A), the paths followed by \rjt and \mdotint are actually disparate. One important difference is the duration of each spectral state phase, as visible in Fig.~\ref{fig:RjMdot1234}. Transitions between hard and soft states ([B,\,C] and [D,\,E]) typically last 15-50 days, whereas other phases can last up to hundreds of days. This is even more visible when focusing on outbursts \one and \three, for example. While [B,\,C] and [D,\,E] form two different pairs separated by 200 days in outburst \one , all four steps are within a 100-day period in outburst \three . These two outbursts are thus very different in their temporal evolutions, and there seems to be no generic behavior.

\subsection{From the HID to the $\dot{m}_{in}-r_J$ plane} \label{sec:paramevol}

One usually assumes that the main parameter controlling the outbursts is the disk accretion rate \citep{Esin97}. It is therefore practical to represent each observational point in a theoretical (parametric) plane: \rj as function of \mdotin .
This is done in Fig.~\ref{fig:RjMdot1234_bis} using the same color-coding and labels (B, C, D, E) as in Fig.~\ref{fig:RjMdot1234}. Each outburst starts in the dark violet phase, follows a path going through points B, C, D, and E, and then finishes in the light yellow phase. These 2D figures are thus physical analogs of the hardness-intensity diagram. Two main comments can be drawn from these figures. First, while the time evolution \rjt and \mdotint of outbursts \one and \three were incompatible (see Sect.~\ref{sec:timevol}), the paths followed in the $\dot{m}_{in}-r_J$ plane show impressive similarities, as illustrated by the similar position of the labels. Second, the pattern drawn in the $\dot{m}_{in}-r_J$ plane is clearer for outburst \four than for the other three. There are two reasons for this. Incidentally, the low-luminosity rising hard phase was not observed, which prevents the rising (dark violet) and decaying (light yellow) phases from obscuring each other. More importantly, the confidence regions of \rj and \mdotin in outburst \four are much smaller in the hard state due to the excellent radio coverage of both the rising and decaying phases (see paper~V). This advocates for the systematic use of radio constraints to derive the physical state of the disk.

We display in the bottom panels of Fig.~\ref{fig:RjMdot1234_bis} the observations for which the radio flux was used to better constrain the fits. All fitting points with huge error bars are now disregarded. For outburst \one we are only left with the last steps of the rising hard state (dark violet) and the very end of the decaying hard state (light yellow). For outbursts \two and \four , the rising and decaying phases are both still visible. For outburst \three a few points at the end of the rising phase and most of the decaying phase are still present. Nevertheless, the track drawn in the $\dot{m}_{in}-r_J$ plane now appears much clearer in the hard state branch, with the least certain fits being ignored.
The main striking point of this representation in the $\dot{m}_{in}-r_J$ plane, where timescales are discarded, is the hint of the existence of a common pattern for \gx (see labels B, C, D, and E). This is discussed further below.

\subsection{A characteristic pattern in the $\dot{m}_{in}-r_J$ plane} \label{sec:rjdemdotin}

\begin{figure}[h!]
  \includegraphics[width=1.0\linewidth]{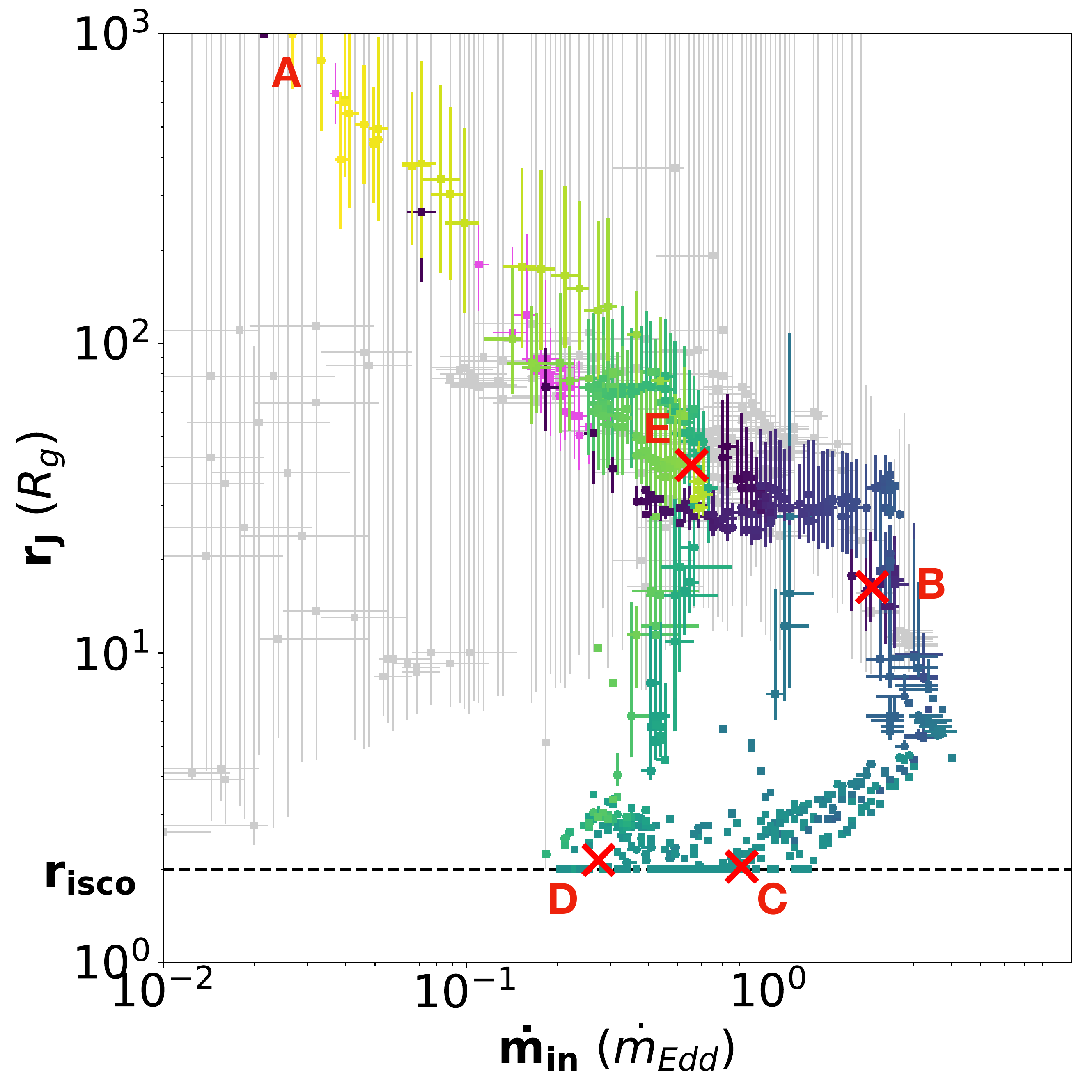}
   \caption{Positions of all observation in the $\dot{m}_{in}-r_J$ plane for the four outbursts of \gx covered by \textit{RXTE}/PCA. All data points are shown in gray, and the constrained observations (see text) are color-coded as in Fig.~\ref{fig:RjMdot1234}. The points in pink are radio-constrained observations of the 2008-2009 failed outburst. Points A-E are placed at the average locations (of the four outbursts) for each phase to illustrate their approximate locations.}
     \label{fig:RjMdot}
\end{figure}

We display in light gray in Fig.~\ref{fig:RjMdot} all observations in the $\dot{m}_{in}-r_J$ plane for the four full outbursts. Here ``constrained observations'' are all observations except those in the hard state with no simultaneous radio coverage (see Sect.~\ref{sec:paramevol}). They represent 852 observations out of the total of 1036 spread over the four full outbursts, and we show them in Fig.~\ref{fig:RjMdot} using the same color-coding as in Figs.~\ref{fig:RjMdot1234} and~\ref{fig:RjMdot1234_bis}. For completeness, the constrained observations from the 2008-2009 failed outburst from \gx \citep[see Fig.~1 in][]{2020A&A...640A..18M} are also shown in pink.

A complete cycle starts in the quiescent state, located at the upper left region of the $\dot{m}_{in}-r_J$ plane; at small $\dot{m}_{in}$ and large transition radius $r_J$, around point A. The start of the outburst corresponds to a simultaneous increase in $\dot{m}_{in}$ and decrease in $r_J$ until point~B, where the system initiates its transition to the soft states. In these intermediate states (from B to C), the spectral evolution is mainly due to a steep decrease in $r_J$, with a much narrower evolution of the disk accretion rate $\dot{m}_{in}$. Soft states are characterized by the non-existence of a JED; the system sticks to the $r_J=r_{isco}$ line, where only an evolution in $\dot{m}_{in}$ is observed. Eventually, \gx exhibits a decrease in $\dot{m}_{in}$ until point~D, where a JED is rebuilt inside-out. Again, these intermediate states mainly correspond to an increase in $r_J$, while the evolution in $\dot{m}_{in}$ is barely visible. At point~E, \gx returns to the hard state branch and then decays back towards the quiescent state in point~A.
Interestingly, we see that the maximum $\dot{m}_{in}$ achieved during the hard state can actually be variable. While outbursts \one, \three , and \four follow the same track (\ie, they have similar $\dot{m}_{in}(B)$), outburst \two reaches a $\dot{m}_{in}(B)$ that is approximately two to three times smaller, $\dot{m}_{in}(B) \simeq 1.0$ instead of $\dot{m}_{in}(B) \simeq 2.5-3.0$. Some degree of freedom must therefore be allowed in the path. Conversely, the lower soft to hard transition (point~D) seems to occur at a similar accretion rate $\dot{m}_{in} \approx 0.2-0.3$. These two results were expected since the upper transitions are known to achieve variable luminosities while the lower ones occur roughly at the same luminosity \citep{Maccarone03, Dunn10}. However, this is now translated into physical quantities for the first time.

While four different complete outbursts are shown in Fig.~\ref{fig:RjMdot}, it clearly exhibits a unique pattern followed by all outbursts when only the constrained observations are used. Moreover, the failed outburst (2008--2009, pink dots) also follows the same track, strengthening the idea of a common path. A failed outburst would be the result of the system not reaching a $\dot{m}_{in}$ high enough (or a \rj small enough) to trigger the transition to a soft state. The unique track suggests a hidden link between \rj and \mdotin (or \mdot ), as rightfully argued in \citet{2019MNRAS.486.2705A}. Specifically, the variations of one parameter would be the answer to the variations of the other, presumably \rj reacting to changes in \mdot due to the long timescales involved. While it has already been predicted in the literature \citep[see, \eg ,][]{Ferreira06, BegelmanArmitage14, Kylafis15}, this is to our knowledge the first demonstration of such a link using an observational fitting procedure \citep[see, however,][for similar studies]{2009MNRAS.396.1415C, Plant15}. We call this path a fingerprint, and we expect all sources to be born with one, but that each will present differences originating from the source's properties, such as the spin of the black hole, or the size or inclination of the system.

\section{Evolution of the disk structure during an outburst} \label{sec:diskstruc}

We now discuss the changes in the thermal structure of the accretion flow during an outburst. Just as in the previous sections, the results presented in this section are model-dependent: the calculations are performed within the JED-SAD paradigm, and thus rely on its assumptions (papers~II and III) and on the chosen fitting procedure (papers~IV).

\subsection{Three routes for energy dissipation} \label{sec:thermalstructure}

The code we developed and used in this study solves for the thermal structure of the hybrid JED-SAD disk configuration. The following local (vertically integrated) equations are solved at each cylindrical radius $r$
\begin{align}
    (1-\delta) \, (q_{acc} - q_{jets}) & = q_{adv, \, i} + q_{ie}, \nonumber \\
    \delta \, (q_{acc} - q_{jets}) & = q_{adv, \, e} - q_{ie} + q_{cool}, \label{eq:2T}
\end{align}
\noindent where we define the total accretion power $q_{acc}$ and the power funneled in the jets $q_{jets}= b(r) q_{acc}$. We assume that the jets take away $30\%$ of the disk accretion power in the JED portion\footnote{Typical values lie in the range $0-80\, \%$ (\ie , $b=0-0.8$), but we chose $b=0.3$ consistently with the other parameters of the JED \citep[][and references therein]{Ferreira97}.} and $0\%$ in the SAD portion (\ie, $b_{JED} = 0.3$ and $b_{SAD} = 0$). As a result, the power funneled in the jets solely depends on the JED accretion power $q_{jets}= 0.3 \, q_{acc,JED}$. Additionally, we assume that electrons and ions share equal parts of the available energy $\delta = 0.5$ \citep[][see their section 2.2]{Yuan14}. We also define the collisional Coulomb heat exchange $q_{ie}$, the radiative cooling term $q_{cool}$, and the advected power transported radially by the ions $q_{adv, \, i}$ and electrons $q_{adv, \, e}$. The local advection term ($q_{adv} = q_{adv, \, i} + q_{adv, \, e}$) can either act as a cooling or a heating term, depending on the sign of the radial derivatives of the internal energy and inflow velocity. We refer the interested reader to section~2 in paper~II.

The global energy budget of our hybrid JED-SAD disk configuration can be obtained by summing the ion and electron energy equations (\ref{eq:2T}), and integrating over all radii $[r_{isco},\, r_{out}]$. This provides 
\begin{align}
    P_{acc} = \frac{G M \dot{M}_{in}}{2 R_{isco}} = P_{jets} + P_{adv} + P_{cool}, 
\end{align}
where $P_\alpha= \int q_\alpha 2\pi r dr$ for each physical term $q_\alpha$ in Eq.~(\ref{eq:2T}). The released accretion power $P_{acc}$ is thus shared between the power carried away by the jets $P_{jets}$, the power advected onto the black hole $P_{adv}$, and the bolometric disk luminosity $L_{bol} = P_{cool}$. It is important to remember that any estimate of these powers is model-dependent and can be subject to important caveats: the observed disk luminosity is only a fraction of $L_{bol}$, the spectrum barely provides any clues on the advected power, and determining the power carried away by the two jets is a substantial task. All estimates in the following are thus done within the JED-SAD framework (\ie, subject to its assumptions and the methods used). To use dimensionless terms, we express these quantities in terms of efficiencies by dividing the above equation by $P_{acc}$: $\eta_{\alpha} = P_{\alpha} / P_{acc}$. These efficiencies verify
\begin{align}
    1 = \eta_{jets} + \eta_{adv} + \eta_{cool}, \label{eq:radeff}
\end{align}
where $\eta_{jets}$ is the global ejection efficiency, $\eta_{adv}$ the global advection efficiency, and $\eta_{cool}$ the global radiative efficiency. When the whole disk is in SAD mode, $\eta_{cool}$ is roughly constant and around unity, regardless of $\dot{m}_{in}$. This case is usually referred to as the radiatively efficient situation, where the disk luminosity varies linearly with the accretion rate $L \propto \dot{m}$ (or \mdotin ). When the inner JED is present, powerful jets are launched, which generates a strong radial torque on the accretion flow. This torque accelerates accretion, allowing for optically thin and geometrically thick solutions, increasing the disk vertical extension and advection processes with it: $\eta_{cool}$ must then necessarily vary with $\dot{m}$ and can eventually reach only a fraction of unity. It is therefore much harder to estimate $P_{acc}$ (and thus \mdot ) directly from the observed spectrum when a JED is involved, unless the behavior of $\eta_{cool}$ is understood (see below).

\begin{figure}[h!]  \centering
  \includegraphics[width=1.0\linewidth]{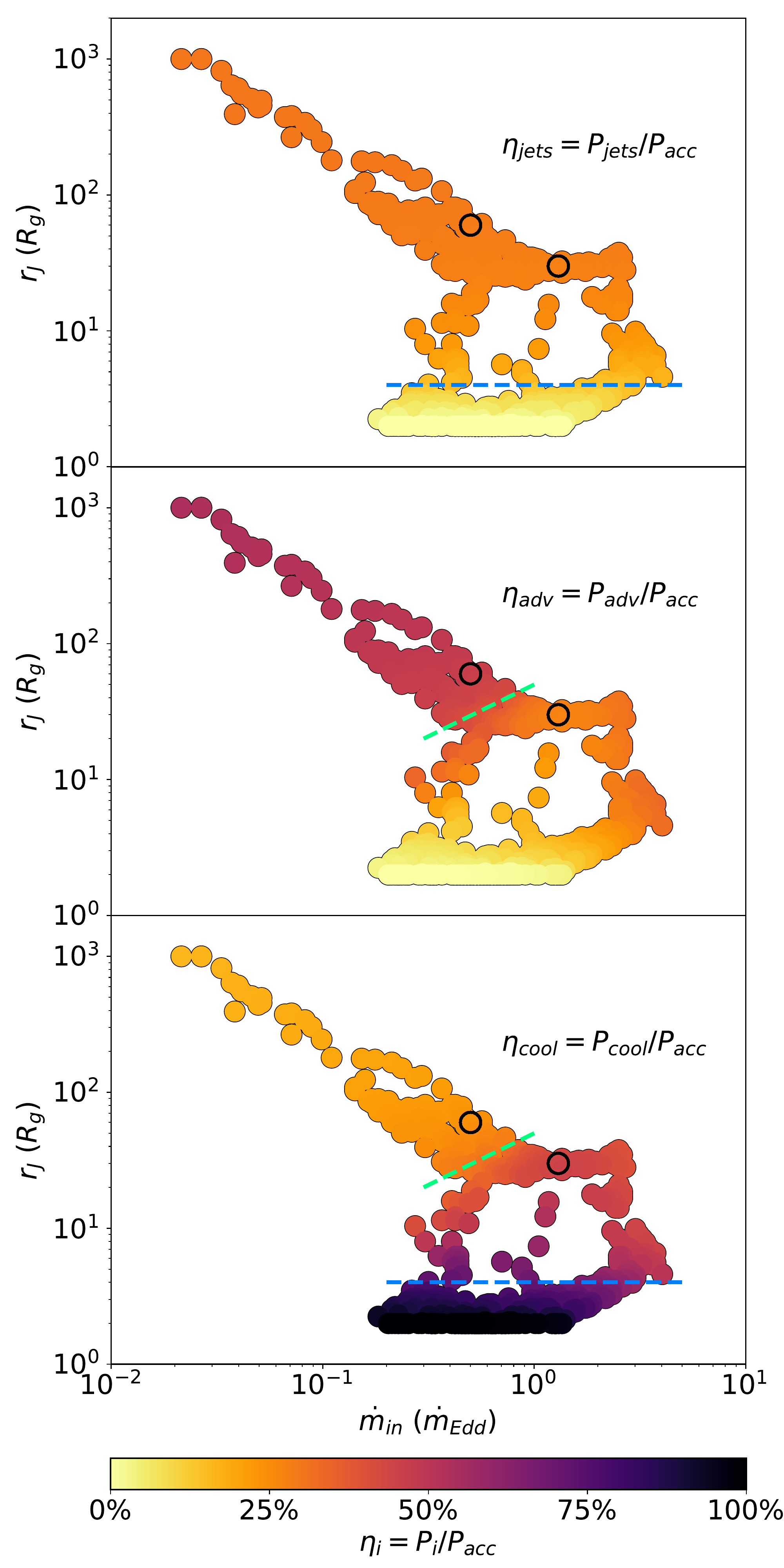} 
   \caption{Positions in the $\dot{m}_{in}-r_J$ plane of all the selected constrained states (see Sect.~\ref{sec:rjdemdotin}), indicating each global efficiency: jets $\eta_{jets} = P_{jets}/P_{acc}$ (top), advection $\eta_{adv} = P_{adv}/P_{acc}$ (middle), and radiation $\eta_{cool} = P_{cool}/P_{acc}$ (bottom). The green line shows states with equipartition (\ie, $\eta_{adv} \approx \eta_{cool} \approx \eta_{jets} \approx 0.30-0.35$), while the blue line separates states where $\eta_{jets}$ is significant (over the line) or not (under the line). We discuss the two solutions circled in black in more detail in Sect.~\ref{sec:thickandslim} and Fig.~\ref{fig:TwoSpectra}.}
  \label{fig:RjMdot_Pi}
\end{figure}

\subsection{Energy dissipation during an outburst} \label{sec:3routes}

We show in Sect.~\ref{sec:fingerprint} that $15$ years of \gx activity results in a characteristic path in the $\dot{m}_{in}-r_J$ plane. For each best fit in this plane we can solve for the thermal structure and derive the various contributions of energy dissipation that reproduce the observed spectrum. We illustrate in Fig.~\ref{fig:RjMdot_Pi} the value of the global ejection efficiency $\eta_{jets}$ (top), advection efficiency $\eta_{adv}$ (middle), and radiative efficiency $\eta_{cool}$ (bottom). For clarity, only the constrained states have been used: hard states with concurrent radio detection, all intermediate states, and all soft states. Equation~(\ref{eq:radeff}) is satisfied by construction at each observational point, and each symbol color varies from light to dark when $\eta_i$ varies from $0$ to $100\%$. This means that each symbol is darker in the figure of its dominating process (see color bar). We recall that all the results derived in this section are done within the JED-SAD framework (\ie, they are model-dependent).

\begin{figure*}[h!]
  \includegraphics[width=1.0\linewidth]{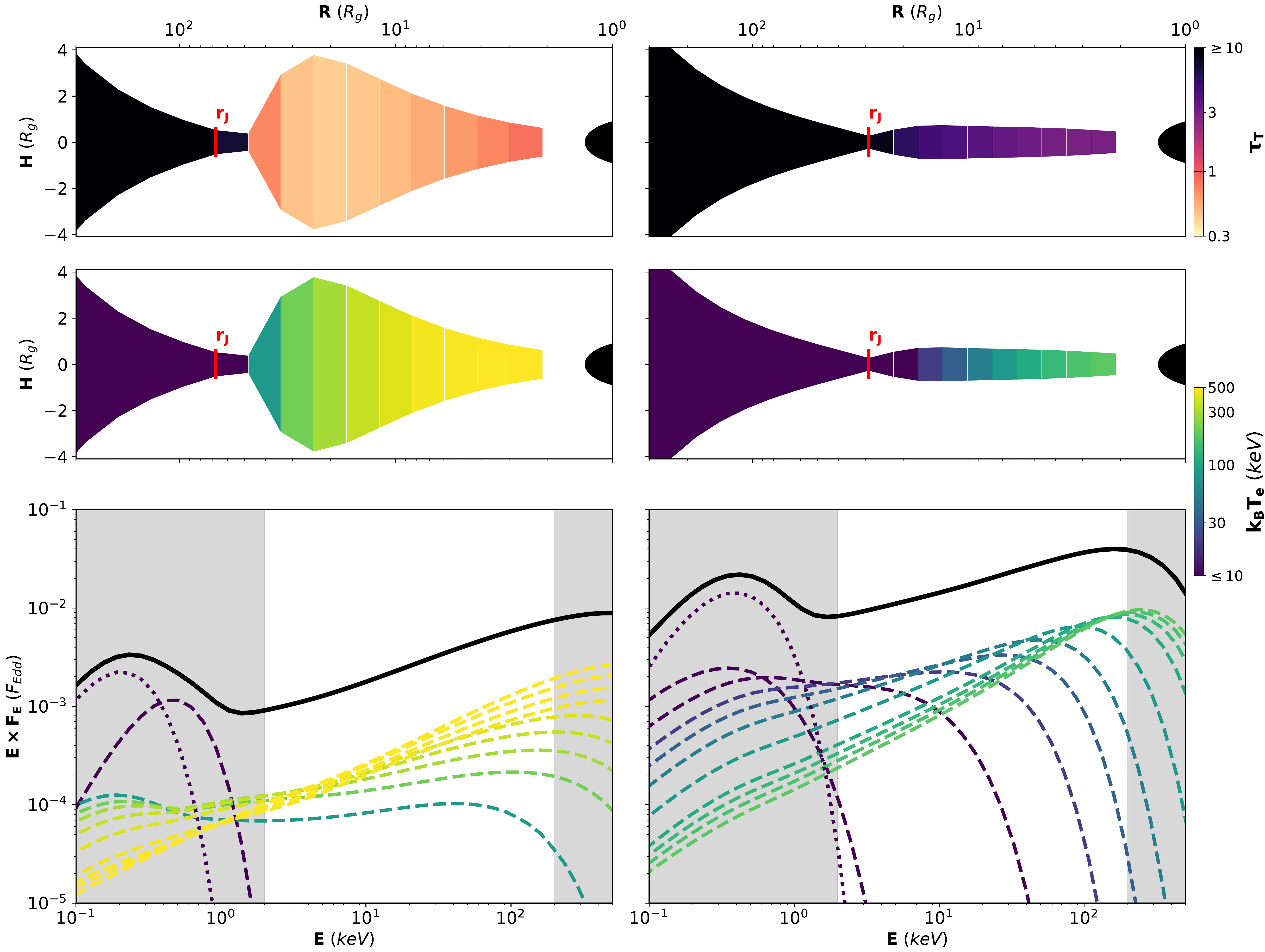}
  \caption{Geometrical shape and spectral energy distribution of the two solutions. Left: JED-SAD solution with $r_J = 65$ and $\dot{m}_{in} = 0.5$. Right: JED-SAD solution with $r_J = 30$ and $\dot{m}_{in} = 1.3$. For both solutions, the radial distribution of the JED is divided into ten portions. Top and middle panels: Radial evolution of the vertical extension of the disk (x-axis at the top) compared to the black hole horizon represented by the black ellipsoid on the right side of each panel. The vertical Thomson optical depth (top) and the electron temperature (middle) are color-coded (top and bottom color bar, respectively). Bottom panels: Total emitted spectrum in black with the contribution from the entire SAD (dotted line) and each JED portion (dashed lines), color-coded according to temperature (same as middle panels). The \textit{RXTE} X-ray range is inside the gray background.}
  \label{fig:TwoSpectra}
\end{figure*}

As expected, the total power carried away by the jets $(\eta_{jets})$ is always near $30\%$ when $r_J \gg r_{isco}$ because we assumed $b = 0.3$ in the JED. A significant decrease is finally visible when $r_J$ reaches $r_J \simeq 2 \, r_{isco} = 4$ (dashed blue line). This is due to the transition to the soft state where very little (but not always zero) energy is channeled in the jets (\ie, $\eta_{jets}$ is at most a few percent). As a result, advection and radiation share only $70\%$ of the released power when $r_J \gg r_{isco}$, and $100\%$ when $r_J = r_{isco}$.
Compared to the ejection efficiency, the advection $(\eta_{adv})$ and radiative $(\eta_{cool})$ efficiencies span a much wider range of values: $\eta_{adv}$ varies between $0\%$ and $55\%$ and $\eta_{cool}$ varies between $15\%$ and $100\%$. In these two panels, the dashed green line is defined as the locus where $\eta_{adv} = \eta_{cool} = 35\%$, namely some sort of equipartition between all energetic processes since $\eta_{jets} = 30\%$. It is important to note that even when advection has the dominant role in the disk structure, it is never greater than $55\%$. These results are thus in strong contrast with the advection-dominated accretion flow \citep[ADAF,][]{NarayanYi95}, presumably because of the difference in the shared energy between the ions and the electrons ($\delta = 0.5$ here, $\delta = 1/2000$ for an ADAF).

On the top left side of the dashed green line, advection is larger than radiation ($P_{adv} > P_{cool}$), whereas on the bottom right side radiative losses overcome the advected energy ($P_{adv} < P_{cool}$). Surprisingly, while all intermediate and soft states are on the same side of the line ($P_{adv} < P_{cool}$), there are hard states on both sides of the line (\ie, with $P_{adv} < P_{cool}$ and with $P_{adv} > P_{cool}$). To understand the reasons and implications of this difference we select two hard state solutions (see the two circles in Fig.~\ref{fig:RjMdot_Pi}), and discuss them in Sect.~\ref{sec:thickandslim}.

\subsection{Thick and slim disk regimes} \label{sec:thickandslim}

We show in Fig.~\ref{fig:TwoSpectra} the thermal structure of the two solutions indicated by black circles in Fig.~\ref{fig:RjMdot_Pi}. We illustrate the radial evolution of their vertical extension on the y-axis of the top and middle panels. The vertical extension of a given accretion flow is usually expected to increase with \mdotin . This is verified here in the \sad region, where in $R = 100 \, R_g$ we have $H \simeq 1 \, R_g$ when $\dot{m}_{in} = 0.5$ (left) and $H \simeq 2 \, R_g$ when $\dot{m}_{in} = 1.3$ (right). In the \jed region, however, the disk is much thinner for the highest \mdotin because less energy is lost through advection or, alternatively, radiative cooling is more efficient. The vertical extension is thus one of the most striking differences between the two regimes: geometrically thick disks $H/R \gtrsim 0.2$ when $P_{adv} > P_{cool}$ (left) and geometrically slim disks $H/R \lesssim 0.1$ when $P_{cool} > P_{adv}$ (right). In comparison, as we have seen, a typical \sad is much thinner with $H/R \approx 0.01-0.02 \ll 1$. We thus label these JED solutions as the ``thick'' regime when $P_{adv} > P_{cool}$ and the ``slim'' regime when $P_{adv} < P_{cool}$. It is important to note that even if we only show two specific solutions here, the properties we discuss apply to all thick and slim states, respectively above and below the dashed green line in Fig.~\ref{fig:RjMdot_Pi}.

In addition to their geometrical differences, there are major differences in the thermal structures of the disks (top and middle panels of Fig.~\ref{fig:TwoSpectra}) and the resulting spectral shape (bottom panel).
In the thick regime (left), the JED is optically thin with $\tau_T \lesssim 0.8 < 1$ (in beige--salmon), and the electron temperature is constant over most\footnote{The first two portions in the JED ($R / R_g \in [65, \, 50]$ and $R / R_g \in [50, \, 35]$) are much colder, with respective temperatures of $k_B T_e \leq 10$\,keV and $k_B T_e \simeq 100$\,keV, because the cold and geometrically thin SAD in $r > r_J$ cools down the JED through radiation and advection processes (see Sect. 2 in paper~III).} of the different \jed annuli: $k_B T_e (r_J) \gtrsim 100-300$\,keV (in light green--yellow). As a result, all sub-spectra have a power-law shape with a high energetic cutoff that is barely detectable in the \textit{RXTE}/PCA spectral band used in this work (bottom panel of Fig.\ref{fig:TwoSpectra}). In this case the total SED, computed by summing all annuli, shares the same properties with a power-law shape whose spectral index arises from the combination of all the sub-power laws. In the slim disk regime (right), the disk is optically slim with $\tau_T \gtrsim 3 > 1$ (purple), and the electron temperature varies radially between $\leq 10$ and $200\,$keV in the JED (dark blue to green). Due to these radial variations of optical depth and temperature, the sub-spectra of the JEDs now have more pronounced spectral differences. For instance, the high-energy cutoff of each sub-spectra appears below $200$\,keV. The sum of all these sub-spectra still gives a power-law shape, but now with a detectable cutoff around $100$\,keV for high-luminosity hard states, consistently with \citet{Motta09}.

There is thus an important transition for the JED within all the hard states observed: from the thick disk regime to the slim disk regime. Thick states are observed at low-luminosity and their energy budgets are $\eta_{adv} \gtrsim 50\%$, $\eta_{cool} \lesssim 20\%$, and $\eta_{jets} = 30\%$. Slim states are observed at high-luminosity and their energy budgets are $\eta_{adv}\simeq 30\%$, $\eta_{cool} \simeq 40\%$, and $\eta_{jets} = 30\%$. This difference is minor in theory since both these regimes are near equipartition with only a slight imbalance between advection and radiation. Moreover, the slopes of both total spectra are quite similar with $\Gamma_{PL} \approx 1.5-2$, showing no clear differences between the two regimes (but see Sect.~\ref{sec:GammaFlux}). However, we show in the following section that the transition between these two regimes has a major observational impact because the radiative efficiency varies from $\lesssim 20\%$ to $40\%$ (\ie, more than a factor of two). Additionally, the spectral shapes at each radius involved are disparate between thick and slim regimes, suggesting that distinct processes could be producing the spectra at different luminosities. This discrepancy could be visible in the evolution of the reflection spectra, in the time-lags observed, and even in the quasi-periodic oscillations. These questions are far beyond the scope of the present paper, although we will see later on that the transition from thick disk to slim disk is directly visible in the variation of the power-law spectral index $\Gamma$ (see Figs.~\ref{fig:Gamma_L110} and \ref{fig:Kolj1}).

\section{Radiative efficiency and the radio--X-ray correlation} \label{sec:secradeff}

\subsection{Radiative efficiency of the accretion flow} \label{sec:radeff}

\begin{figure}[h!]
  \centering
  \includegraphics[width=1.0\linewidth]{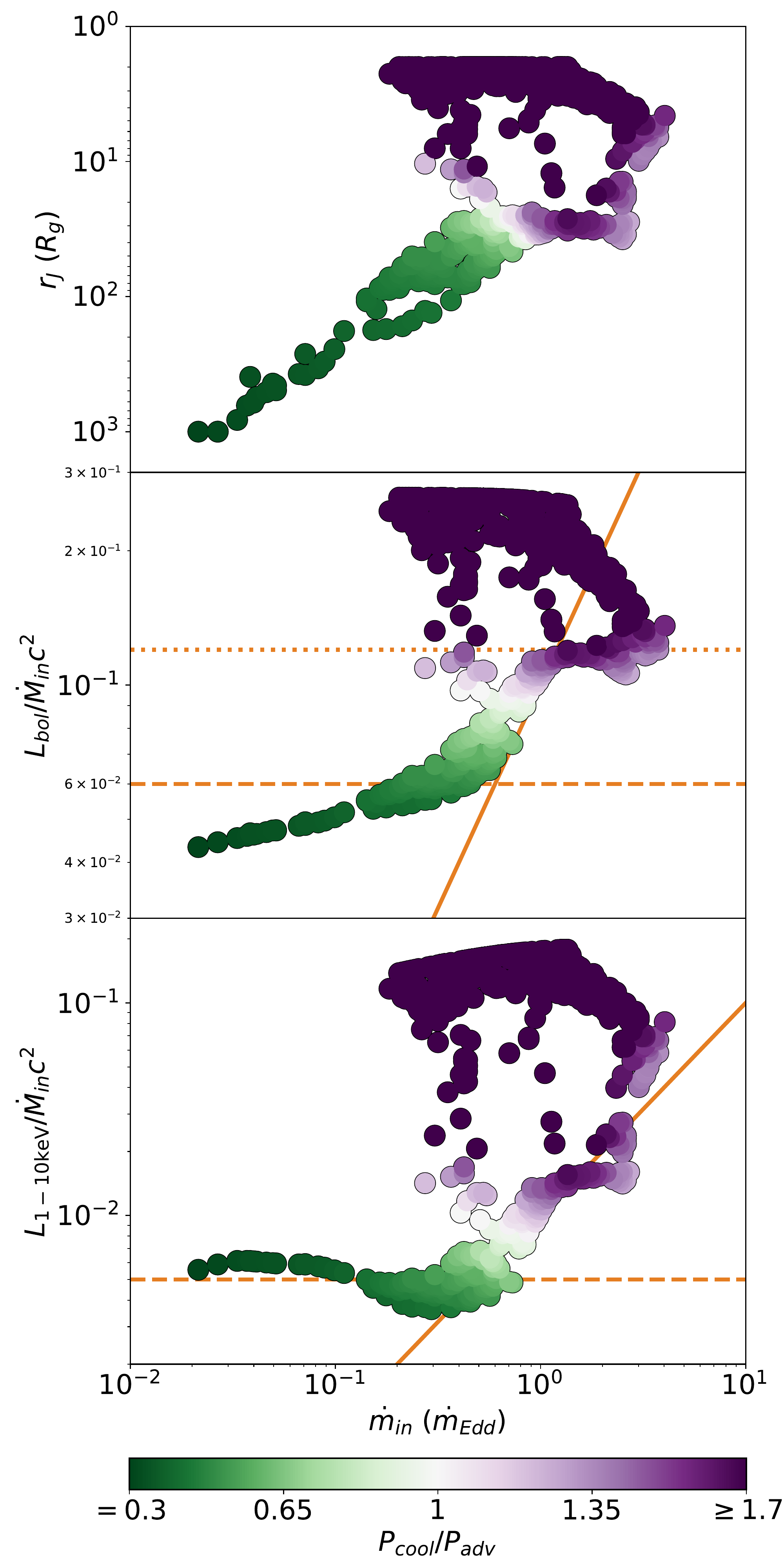} 
   \caption{Observations of \gx in the $\dot{m}_{in}-r_J$ plane (top), bolometric luminosity ($L_{\mathrm{bol}}$) computed from our model (middle), and $1-10\,$keV luminosity ($L_{1-10\,\mathrm{keV}}$) as function of \mdotin (bottom). In all panels the color is the ratio of bolometric radiated power to the power advected (see color bar). The different lines are shown in orange to illustrate different regimes: solid when $L \propto \dot{m}_{in}^2$, and dotted and dashed for the two different regimes when $L \propto \dot{m}_{in}$.}
 \label{fig:L_vs_mdotin}
\end{figure}

We now focus on the disk radiative efficiency. For each observation, we show in Fig.~\ref{fig:L_vs_mdotin} the evolution of \rj (top), the bolometric \radeff (middle), and the unabsorbed X-ray \radeff ($1-10$\,keV, bottom) as function of the disk inner accretion rate \mdotin . Both \radeffs are defined as the ratio of their associated luminosity with respect to the total available power $\dot{M}_{in} c^2$, labeled $\varepsilon_{\mathrm{bol}} = L_{\mathrm{bol}} / (\dot{M}_{in} c^2)$ and $\varepsilon_{1-10} = L_{1-10\,\mathrm{keV}} / (\dot{M}_{in} c^2)$ in all the following. The bolometric radiative efficiency can be linked to the cooling efficiency with $\varepsilon_{\mathrm{bol}} = \eta_{cool} \eta_{acc}$, where $\eta_{acc} = 1/(2 r_{isco}) = 1/4$ is the accretion efficiency of a black hole of spin $0.93$. In other words, an accretion flow radiating $100 \, \%$ of its accretion power will only emit $25 \, \%$ of its total available power (\ie, $25 \, \% \, \dot{M}_{in} c^2$), while a flow radiating $10 \, \%$ of its accretion power will emit $2.5 \, \% \, \dot{M}_{in} c^2$.
We note that the top panel of Fig.~\ref{fig:L_vs_mdotin} is the same as the bottom panel of Fig.~\ref{fig:RjMdot_Pi}, but with an inverted y-axis and using different color maps and scales: the color now indicates the ratio of advected to radiated powers. This allows us to highlight the two different JED regimes discussed in Sect.~\ref{sec:3routes}: thick in green, slim in purple, as well as their transition in white. The color bar includes the state with minimum radiative power ($P_{cool} \approx 0.3 \, P_{adv}$), whereas the maximum of the color bar is set to $P_{cool} = 1.7 \, P_{adv}$, even if it reaches $P_{cool} \approx 10^4 \, P_{adv}$ in the soft states where only a SAD is present. {We again recall that the results derived in this section are done within the JED-SAD framework (\ie, they are model-dependent).}

These different portions of the outburst are directly visible in the evolution of the bolometric radiative efficiency $\varepsilon_{\mathrm{bol}} = L_{\mathrm{bol}} / (\dot{M}_{in} c^2)$ as a function of \mdotin (middle panel). This figure is a slight modification of the $\dot{m}_{in}-r_J$ plane because the thermal structure at a given \mdotin strongly depends on \rj . However, variations in \mdotin produce a more complicated picture, especially at the transition between the two states (white). At low \mdotin , we are in the thick disk regime (green) with $L_{\mathrm{bol}} \lesssim 0.06 \, \dot{M}_{in} c^2$ (\ie, $\varepsilon_{\mathrm{bol}} \lesssim 6\%$; dashed orange line). As \mdotin increases, we transition to the slim disk regime (violet and purple) where $L_{\mathrm{bol}} > 0.12 \, \dot{M}_{in} c^2$ (\ie, $\varepsilon_{\mathrm{bol}} > 12\%$; dotted orange line). Interestingly, the transition from thick to slim (in white) follows an $\varepsilon_{\mathrm{bol}} \propto \dot{m}_{in}$ track showed in solid orange. A regime with $\varepsilon_{\mathrm{bol}} \propto \dot{m}_{in}$ (\ie, $L_{\mathrm{bol}} \propto \dot{M}_{in}^2$) has been labeled as radiatively inefficient in the past \citep[see, e.g.,][section 4.3.3]{Coriat11}. In our case, however, there is no such regime. The portion with $L_{\mathrm{bol}} \propto \dot{M}_{in}^2$ is only a transition between two different states of different radiative efficiencies, both of which are actually radiatively efficient (as opposed to the flows that are usually considered radiatively inefficient, see below). As \rj decreases to \risco , the \radeff increases because the \jed radial extension shrinks to give more ground to the \sad . We finally reach a maximum value of $L_{\mathrm{bol}} \approx 0.25 \, \dot{M}_{in} c^2$ (\ie, $\varepsilon_{\mathrm{bol}} \approx 25\%$ when only a \sad is present.

In the bottom panel of Fig.~\ref{fig:L_vs_mdotin}, the evolution with the accretion rate is altered by the energy band chosen (see paper III, section 4.2). The thick disk regime (in green), where bolometric luminosity increased moderately with \mdotin , now follows a nearly constant efficiency when only looking at the $1-10$\,keV range. This gives rise to a first phase with constant $\varepsilon_{1-10}$ (\ie, $L_{1-10\,\mathrm{keV}} \propto \dot{m}_{in}$; orange dashed line). However, the phase $L_{\mathrm{bol}} \propto \dot{m}_{in}$ observed during the slim disk regime (violet) has disappeared. Instead, the transition from thick to slim seems to merge with the slim disk regime itself. This gives birth to a second phase with a surprisingly steep slope $\varepsilon_{1-10} \propto \dot{m}_{in}$ (\ie, $L_{1-10\,\mathrm{keV}} \propto \dot{m}_{in}^{2}$; orange solid line). One usually associates this phase to a radiatively inefficient accretion flow, but around $40\%$ of the accretion power is still radiated away in our model. Both phases (dashed and solid lines) should instead be considered as radiatively efficient, with around $\varepsilon_{\mathrm{bol}} \approx 6\%$ to $12\%$ of the total available power\footnote{Not to be confused with the accretion power $P_{acc}$, different by a factor $\eta_{acc} = 1/(2 r_{isco}) = 0.25$ here.} radiated away, and $\varepsilon_{1-10} \approx 0.5\%$ to $3\%$ in the $1-10\,$keV band. However, we can compare these regimes to an even more efficient regime during the soft-state $r_J = r_{isco}$ (dark violet), where all the accretion power is radiated away (\ie, $\varepsilon_{\mathrm{bol}} \approx 25 \%$) with $\varepsilon_{1-10} \approx 10-15 \%$ in the soft X-ray band.

These results illustrate that the presence of two regimes and the difference in their \radeff values generates different slopes in the $L-\dot{M}$ evolution. One should thus be very careful when discussing \radeffs . Accretion rate changes lead to structural changes that can lead to misconceptions or even misinterpretations when filtered through a given energy range. While we do observe a $L \propto \dot{M}_{in}^2$ phase during the outburst, this phase is never associated with what is usually called a radiatively inefficient accretion flow since the flow always radiates more than $20\%$ of its available accretion power (about $5\%$ of the total available power).

\subsection{Evolution of the power-law index} \label{sec:GammaFlux}

In the previous sections of this paper we illustrate the change in the thermal structure of the accretion flow during the hard state. More precisely, we show that there are two different regimes depending on the thermal structure of the disk: thick disk when $P_{adv} > P_{cool}$ and slim disk when $P_{adv} < P_{cool}$. However, all the changes presented are model dependent: all calculations are performed within the JED-SAD paradigm, and thus rely on its assumptions (papers~II and III) and on the chosen fitting procedure (papers~IV). Such a transition has already been observed using model-independent estimates: the hardness ratio or the spectral index of the power law \citep[see, e.g.,][]{2011MNRAS.417..280S}. In this work we decided to use the spectral index because any hardness ratio requires a choice of two spectral bands and we wanted to remain as generic as possible.

\begin{figure}[h!]
  \includegraphics[width=1.0\linewidth]{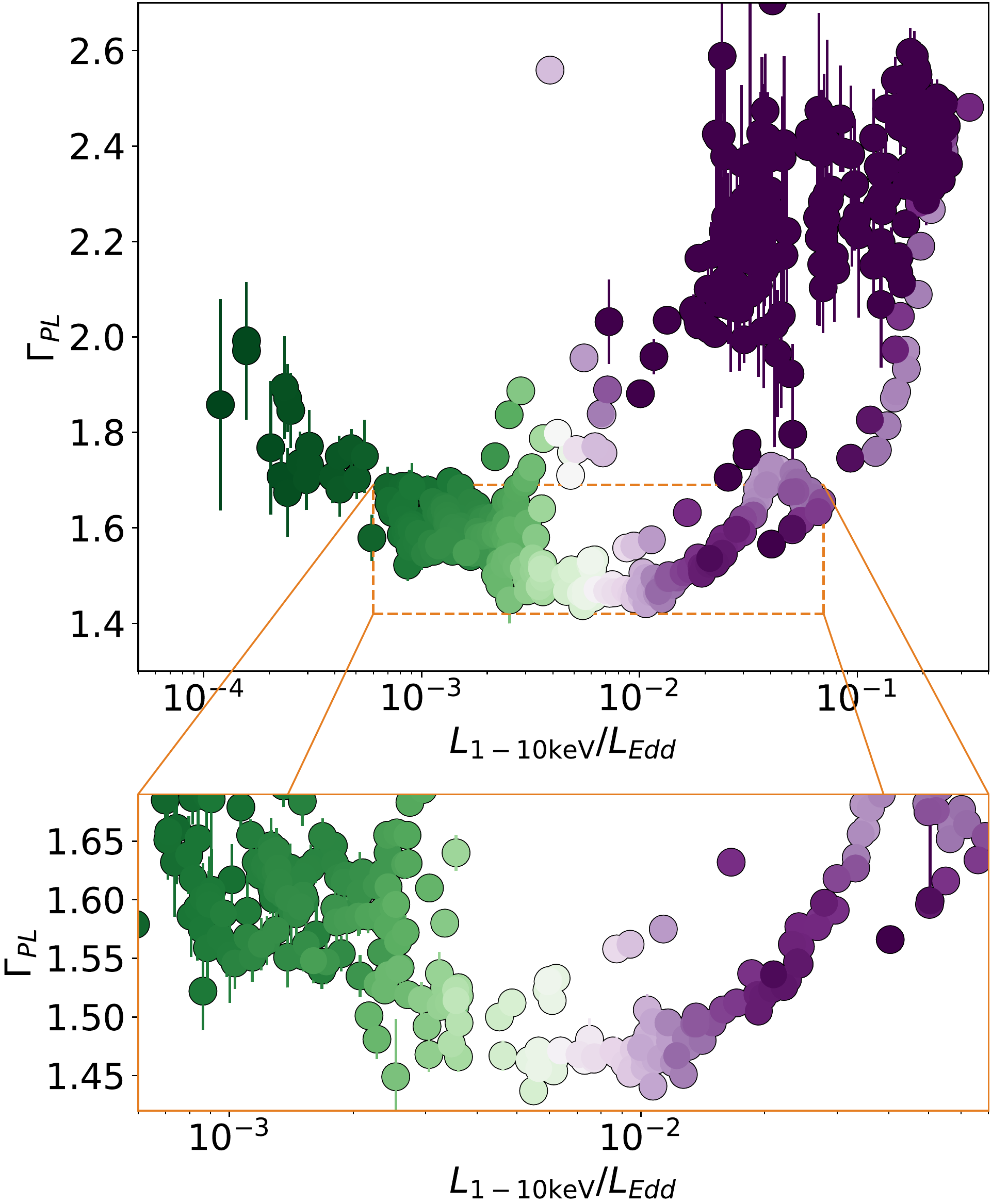}
  \caption{Evolution of the power-law index $\Gamma_{PL}$ of each spectrum \citep{Clavel16} as a function of its luminosity in the $1-10$\,keV range. All four outbursts during the 2000s are overplotted (color-coding as in Fig.~\ref{fig:L_vs_mdotin}). The bottom panel is a zoom-in on the zone inside the orange dashed rectangle in the top panel. For clarity, we do not show observations when $\Gamma_{PL}$ is unconstrained (soft states), but the usual values lie in the $\Gamma_{PL} \approx 2.0-3.0$ range.}
  \label{fig:Gamma_L110}
\end{figure}

We show in Fig.~\ref{fig:Gamma_L110} the evolution of the spectral index of the power law from \citet{Clavel16}. We show all observations during the four outbursts except when $\Gamma_{PL}$ is unconstrained in the soft state. The top panel shows the entire range of values, while the bottom panel shows a zoom-in on the transition between thick and slim regimes (see below). In this figure an outburst starts on the left side with $\Gamma_{PL} \approx 1.8-2.0$ and $L_{1-10\,\mathrm{keV}} \approx 10^{-4} \, L_{Edd}$, and then runs through the cycle counterclockwise. All hard states are roughly found when $\Gamma_{PL} \leq 1.8-2.0$, while soft states are around $\Gamma_{PL} \approx 2.2-2.6$. The same color map as in Fig.~\ref{fig:L_vs_mdotin} is used here, illustrating the transition from thick disk solutions ($P_{cool} < P_{adv}$, in green) to slim disk solutions ($P_{cool} > P_{adv}$, in purple). As already seen in Fig.~\ref{fig:L_vs_mdotin} (bottom panel), the transition (in white) from the thick to the slim solution happens around $L_{1-10\,\mathrm{keV}} \approx 5 \times 10^{-3} - 10^{-2} \, L_{Edd}$. What is interesting here is that the transition is also concurrent with a change in the evolution of $\Gamma_{PL}$ (see zoomed-in portion in bottom panel). When $L_{1-10\,\mathrm{keV}} \leq 5 \times 10^{-3} \, L_{Edd}$, $\Gamma_{PL}$ decreases with luminosity, but when $L_{1-10\,\mathrm{keV}} \geq 10^{-2} \, L_{Edd}$, $\Gamma_{PL}$ increases with luminosity.

There is thus an important change in the spectral shape around\footnote{The mass of \gx is not well constrained, and this luminosity could actually lie anywhere in the range $L_{1-10\,\mathrm{keV}} \approx 10^{-3}-10^{-2} \, L_{Edd}$.} $L_{1-10\,\mathrm{keV}} \simeq 5 \times 10^{-3} \, L_{Edd}$. The evolution of $\Gamma_{PL}$ is fully consistent with the locus of the thick to slim transition expected from our JED-SAD framework. The evolution in $\Gamma_{PL}$ in the hard state appears then as a convenient tool to trace the change in the accretion flow structure.

\subsection{Evolution of the spectral shape during the hard state} \label{sec:HS_obs1}

\begin{figure}[h!]
  \includegraphics[width=1.0\linewidth]{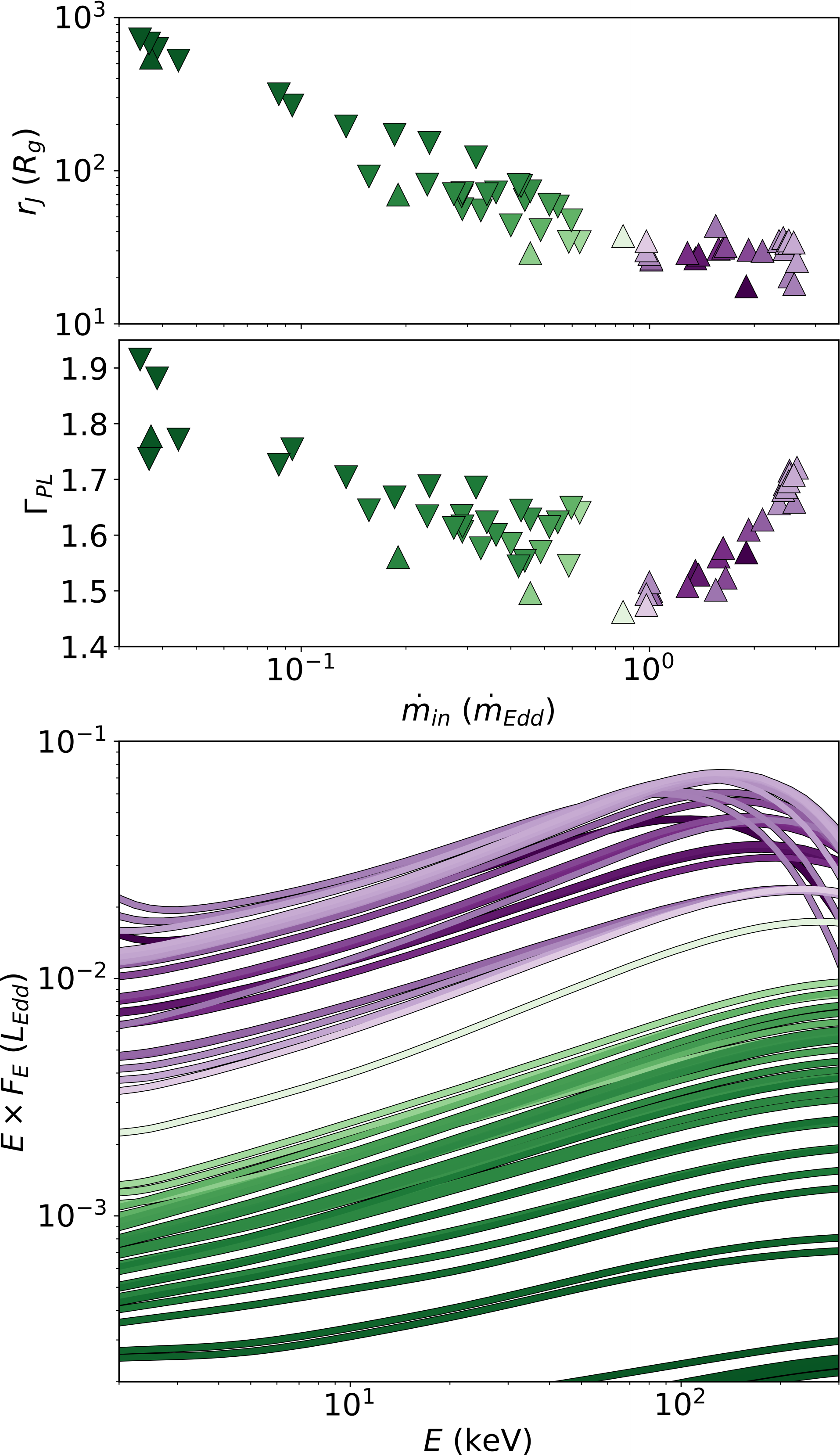}
  \caption{Subset of all hard state observations with detected radio fluxes, using the same color-coding as in Fig.~\ref{fig:L_vs_mdotin}. The symbol shapes show the rising (up-pointing triangle) and decaying (down-pointing triangle) hard state. Top: Distribution of these observations in the $\dot{m}_{in}-r_J$ plane (x-axis at top). Middle: Spectral index of the power law $\Gamma_{PL}$ from the \citet{Clavel16} as function of the \mdotin from our fits. Bottom: Theoretical spectra associated with each solution \pair in the (theoretical) \textit{RXTE} range $2-300$\,keV.}
  \label{fig:Kolj1}
\end{figure}

In this section we consider only the constrained hard states (\ie, the hard states when radio flux has been observed). These states are all fit with $r_J \gg r_{isco}$ (\ie, with $30\%$ of the power channeled in the jets and $70\%$ shared between $P_{adv}$ and $P_{cool}$). We show in Fig.~\ref{fig:Kolj1} the $\dot{m}_{in}-r_J$ plane (top), the power-law spectral index $\Gamma_{PL}$ (middle), and the associated best fit JED-SAD spectra in the $2-300$\,keV range (bottom) for each observation. The color scale is as in Fig.~\ref{fig:L_vs_mdotin}: thick disk spectra in green $(P_{cool} < P_{adv})$, slim disk spectra in purple $(P_{cool} > P_{adv})$, and the transition (equipartition $P_{cool} = P_{adv}$) in white.

We see in the $\dot{m}_{in}-r_J$ plane (top panel) that the transition between the thick and the slim regimes is barely visible; without the colors we could not locate the transition on this panel, around $r_J \simeq 30$ and $\dot{m}_{in} \simeq 0.75$. As seen in Sect.~\ref{sec:radeff}, the luminosity evolves as $L \propto \dot{m}_{in}^2$ at the transition between the two regimes. A slight increase in $\dot{m}_{in}$ then translates into a dramatic rise in luminosity: we should expect the transition between these regimes to generate a gap in luminosity. This gap is indeed observed in the hard state spectral evolution\footnote{The absence of observations at that luminosity could also be a result of the observational strategy. For example, outburst \four was first detected at a luminosity higher than this gap.}, as seen in the bottom panel of Fig.~\ref{fig:Kolj1} and already discussed in \citet{Koljonen19}. In our model the observed gap simply originates from a change in the thermal structure of the accretion flow, and eventually in \radeff .

As discussed in Sect.~\ref{sec:GammaFlux}, while the change in \radeff during an outburst is model-dependent, the spectral index of the power law $\Gamma_{PL}$ is not. We show in the middle panel of Fig.~\ref{fig:Kolj1} the evolution of $\Gamma_{PL}$ as a function of \mdotin for all our constrained hard states\footnote{We show $\Gamma_{PL}$ as a function of \mdotin to better compare with the \rjdemdotin curve, but a similar shape is naturally recovered when represented as a function of the X-ray luminosity (see bottom panel of Fig.~\ref{fig:Gamma_L110}).}. We see in this figure that the gap in luminosity (in white) coincides with a change in the evolution of $\Gamma_{PL}$. At low accretion rates (or luminosity, see Fig.~\ref{fig:Gamma_L110}), the power-law spectral index decreases as \mdotin increases. At high accretion rates, however, the spectral index now clearly increases with \mdotin . There are thus two different groups of hard state spectra, corresponding to the two different regimes discussed in Sect.~\ref{sec:3routes}: thick disk and slim disk. The spectral shapes are very similar between the two regimes and mainly differ by their fluxes. As a result, finding this transition can be tricky. We illustrate in this work that one can do so either by using our JED-SAD model, or simply by tracking the changes in $\Gamma_{PL}$ \citep[see, e.g.,][]{2011MNRAS.417..280S}. Because we believe the physical structure of the accretion flow to be similar in all X-ray binaries, we expect this transition to be present for all objects around a similar luminosity (\ie, around $L_{1-10\,\mathrm{keV}} \approx 10^{-3} - 10^{-2} \, L_{Edd}$). Moreover, we expect this transition to have an impact on the radio--X-ray correlation, as discussed in Sect.~\ref{sec:radioXray}.


\subsection{The radio--X-ray correlation} \label{sec:radioXray} 

\begin{figure}[h!]
  \includegraphics[width=1.0\linewidth]{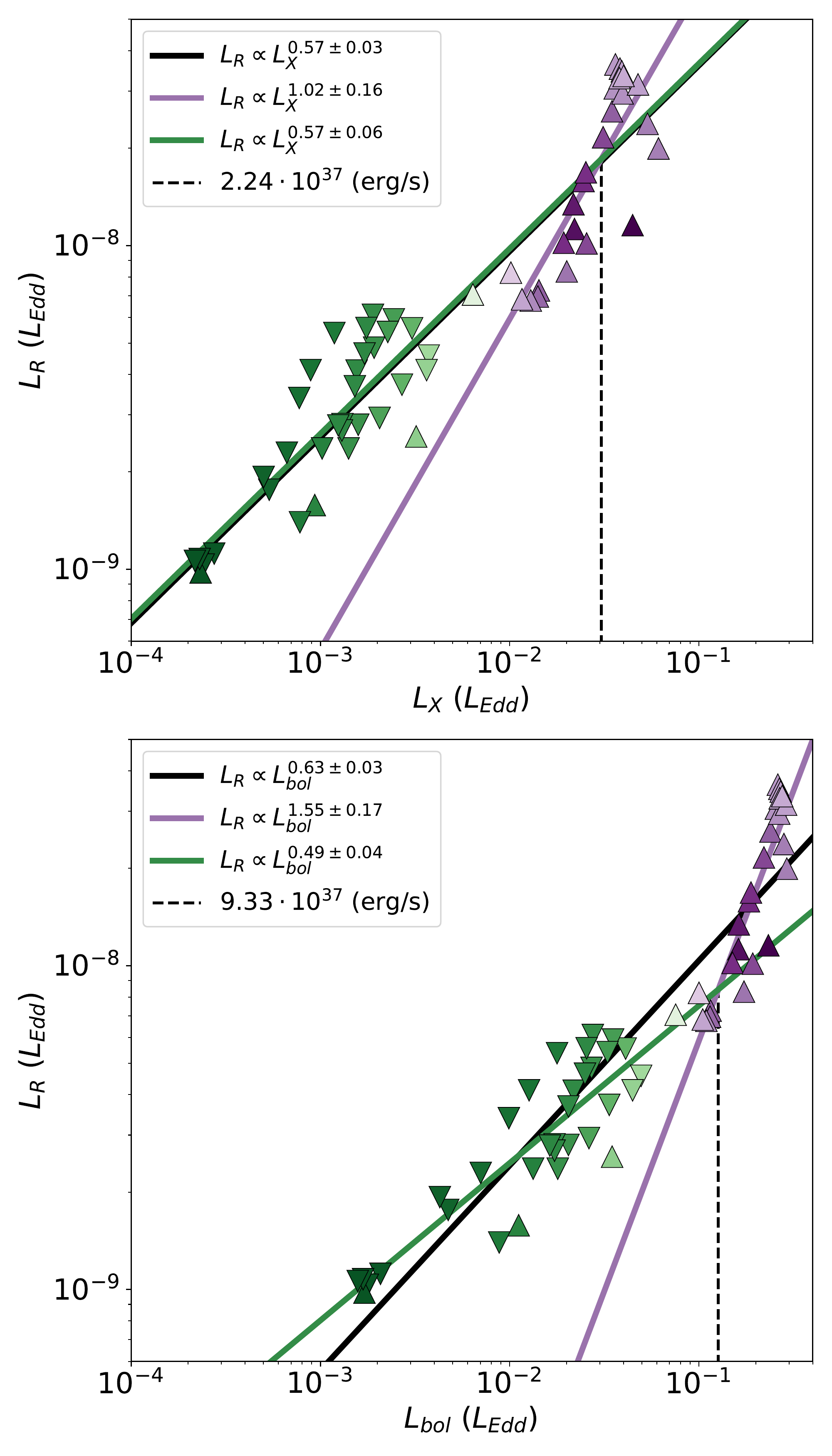}
  \caption{Correlations between the observed radio fluxes and the fitted X-ray flux emitted by the accretion flow in two different energy ranges: classical $1-10\,$keV range (top) and bolometric (bottom). In each panel the symbol shows a rising (up-pointing triangle) or decaying (down-pointing triangle) hard state, and the color is the ratio $P_{cool}/P_{adv}$ of the radiative to the advected power (see color bar in Fig.~\ref{fig:Kolj1}). The solid lines indicate our fits: global correlations (black), with only $P_{adv}>P_{cool}$ solutions (green) and with only $P_{adv} < P_{cool}$ solutions (purple). A dashed line illustrates the location of the crossing point of the thick disk and slim disk correlations.}
  \label{fig:Kolj3}
\end{figure}

In this section we only isolate the constrained hard states as in Sect.~\ref{sec:HS_obs1}, and we focus on the impact of the \radeff regimes on the radio--X-ray correlation $L_R \propto L_X^a$. In this correlation, the luminosity $L_R$ is derived from the observed radio flux densities, usually around $5-9$\,GHz, and $L_X$ from the flux observed in either the $1-10\,$keV or the $3-9\,$keV energy range \citep[see, e.g.,][]{Coriat11, Corbel13}. We use the $8.6-9.0$\,GHz radio band and the $1-10\,$keV X-ray range to be consistent with the known correlation from \citet{Corbel13}. We show in Fig.~\ref{fig:Kolj3} the observed radio luminosity $L_R$ as a function of the observed $1-10\,$keV (top) and bolometric (bottom) X-ray luminosities. We use the same color-code as in Fig.~\ref{fig:L_vs_mdotin} (\ie, thick hard states in green, slim hard states in purple, transition in white). We recall that the top panel of Fig.~\ref{fig:Kolj3} is purely observed data, while the bottom panel shows bolometric fluxes that have been obtained using our model. As a result, the top panel is model-independent, but the bottom panel is not and it relies on our assumptions and fitting procedure.

When we fit all the observations together in the $1-10$\,keV range (black line) we retrieve the usual correlation with $a = 0.57 \pm 0.03 \simeq 0.6$ \citep{Corbel03, Gallo03, Coriat11, Corbel13}. This was expected since we use the same X-ray ranges as \citet{Corbel13}.
However, things become interesting when the two different regimes are considered independently. While we obtain a similar fit using only the thick disk observations ($a = 0.57 \pm 0.06$), the correlation becomes much steeper in the case of slim disk observations ($a = 1.02 \pm 0.16$). We believe that the reason for this discrepancy is the sensitivity of the disk radiative efficiency $\varepsilon_{1-10}$ to changes in $\dot{m}_{in}$, as explained in Sect.~\ref{sec:radeff}. When the system transitions from the thick to the slim regimes both \rj and \mdotin undergo steady changes. While the X-ray emission shows swift changes in \radeff near the equipartition $P_{cool} = P_{adv}$ zone (white), the jet radio emission undergoes slow and steady changes: there is no apparent reason for its \radeff to vary as well. This produces the observed plateau in the $L_R-L_X$ curve around $L_R = 7 \times 10^{-9} \, L_{Edd}$, providing a natural explanation for the wiggles seen in the radio--X-ray correlation curve for \gx . We note, however, that the locus of the wiggles corresponds to the soft to hard transition (\ie, the rebuilding of the jets). It is thus possible that these wiggles are a result of the jet building, as is discussed in \citet{2022A&A...657A..11B}.

For completeness, we show in the bottom panel of Fig.~\ref{fig:Kolj3} the radio luminosity as a function of the bolometric luminosity. We recall here that the bolometric luminosity is a result of our physical model, not an extrapolation of the disk and power-law shapes. When including all observations with detected radio and X-rays, we find a similar correlation with $a = 0.63 \pm 0.03$. However, there is now an even clearer difference between the green and the purple states, the former providing a correlation with $a = 0.49 \pm 0.04$ and the latter a much steeper exponent $a = 1.55 \pm 0.17$.
This result shows that these wiggles are not an effect of the spectral energy range used for observations and should always be seen. However, the critical luminosity $L_c$ where the two correlations meet, namely where a break with the low-luminosity branch becomes distinguishable, does depend on the energy band. For \gx and the set of dynamical parameters used, the critical luminosity in the $1-10$ keV energy range is $L_c \simeq 3 \times 10^{-2}\,L_{Edd} \simeq 2 \times 10^{37}\,\text{erg/s}$ and becomes $L_c \simeq 12 \times 10^{-2} \, L_{Edd} \simeq 9 \times 10^{37}\,\text{erg/s}$ in bolometric.
As said above, these luminosities happen to be close to where the actual hysteresis cycle starts, namely the point in the HID where the rising hard state branch meets the decaying horizontal soft-to-hard branch. As a result, all hard states with $L_X > L_c$ were observed during the rising phase (up-pointing triangles in Figs.~\ref{fig:Kolj1} and \ref{fig:Kolj3}) whereas most (but not all) of the hard states with $L_X < L_c$ were observed during the decaying state (down-pointing triangles). This raises questions about the possible differences between rising and decaying phases \citep{2018MNRAS.481.4513I, 2022A&A...657A..11B}. However, our modeling shows that this may only be due to the evolution of the JED radiative efficiency as the disk accretion rate increases.

\subsection{The existence of outliers}

An important aspect of the radio--X-ray correlation is the presence of outliers; namely X-ray sources that do not follow the standard correlation, but a rather steeper correlation (at least at sufficiently high flux) with $L_R \propto L^{\approx 1.4}$ \citep[][]{Coriat11, 2012MNRAS.423..590G, Corbel13}.
In the present study we show that we retrieve the standard correlation followed by \gx when all hard states are used. However, we find $L_R \propto L^{\approx 1-1.5}$ when considering only the slim hard states. This result suggests that \gx would behave similarly to H 1743-322, where two different tracks $L_R \propto L^{\leq 0.6}$ and $L_R \propto L^{\geq 1.0}$ are observed \citep[][Figure~9]{Corbel13}. Interestingly, most outliers seem to be observed at high luminosities ($L_{1-10 \, \mathrm{keV}} \geq 10^{36} \, \mathrm{erg.s}^{-1}$) (\ie, where the inner \jed becomes slim). This raises the question of the existence of outliers: Do {outliers} exist, or is the observed sample only located in this slim disk regime?

To answer these questions, similar studies on more outbursts from both \gx and from other sources would need to be performed. Such a study should also investigate the uniqueness of the transition luminosity. This transition in the disk radiative efficiency, seen as a plateau in the $L_R-L_X$ correlation, is inherent to our JED-SAD model (at least for the parameters used for \gx).
If such a transition is not observed in other objects then the change in the slope observed in outliers cannot arise from changes in the disk radiative efficiency. This would mean that another physical factor must be at work, most probably related to the jet emission itself: collimation properties, internal shock conditions \citep{2013MNRAS.429L..20M, 2014MNRAS.443..299M, 2020MNRAS.498.3351M}, or even the interplay between the \citeauthor{BP82} jet emitted from the JED and the inner spine emitted from the black hole ergosphere \citep{BZ77}. This opens quite interesting prospects, far beyond the scope of the present study.

\section{Caveats} \label{sec:caveatsradio}

Our JED-SAD modelling comes with some caveats. These caveats were thoroughly discussed in the previous papers in this series, but we recall here the most important points.

We perform our calculations using a Newtonian potential and a non-relativistic version of the radiative transfer code \textsc{Belm} \citep{Belmont08, Belmont09}. As the disk material plunges into the black hole, a smaller fraction of the mechanical energy is actually released: the \citet{NT73} turbulent dissipation is smaller than in the \citet{SS73} non-relativistic disk (see \citealt{1974ApJ...191..499P} and also discussion in \citealt{2012MNRAS.420..684P}). We thus expect our model to underestimate the disk accretion rate \mdotin for radii $r_J \leq 6$. This approximation is not crucial for large transition radii $r_J \gg r_{isco}$ (\ie, for the states thoroughly studied in this paper), but it can become important when discussing the shape of the track followed in the $\dot{m}_{in}-r_J$ plane at small \rj . This will be discussed in a forthcoming work.

The calculations performed in this paper only use the $3-25$\,keV \textit{RXTE}/PCA data, in which the high-energy cutoffs cannot be constrained. Because they are unreliable in this range, the cutoffs seen in Fig.~\ref{fig:Kolj1} have not been implemented in the fitting procedure. However, their presence is inevitable within the JED-SAD paradigm. These cutoffs were not necessary in the fits, but they are unavoidable in the resulting spectra. This can be tested in broadband ($>10$\,keV) data sets, as is done in the case of Swift+NICER+NuSTAR observations from the black hole X-ray binary MAXI J1820+070 \citep{2021A&A...656A..63M}. This may also explain the possible differences with the real observed cutoff that can be seen in Figure~2 of \citet{Koljonen19}.

In this work we use the equation that was solved in \citet{Marcel18a} to separate all energy components: $P_{acc} = P_{jets} + P_{adv} + P_{cool}$. However, it has recently been argued that winds could be present at all times during an outburst (see, e.g., \citealt{2015ApJ...814...87M}, and also \citealt{2021A&A...649A.128P}). A more complete equation should thus be $P_{acc} = P_{jets} + P_{adv} + P_{cool} + P_{winds}$ at each radius, and with $P_{jets} = 0$ or $P_{winds} = 0$ depending on whether jets and/or winds are produced. In our view, winds would be produced in the SAD portion, hence not affecting the JED. We have thus decided to neglect the power carried away by the winds at all times (\ie, $P_{winds} = 0$). On a similar note, all observed soft states display non-thermal tails generally observed above a few keV: the hard tail \citep{McConnell02, Remillard06}. Within our framework, this hard component is thought to be related to coronal dissipation and the production of a non-thermal electron population \citep[see, \eg ,][]{Galeev79, Gierlinski99}. This non-thermal population is not taken into account, however, and our treatment of the hard tail is parametric (see discussion and Fig.~5 in paper~III). This may also lead to some uncertainties on the energy equation during the soft states (\ie, not relevant in the \radeff and radio--X-ray correlation we discussed). Moreover, non-thermal effects could also have an impact on the disk structure and its equilibrium during any given state \citep{McConnell02, 2009MNRAS.392..570M}, but we chose to ignore these effects in this work.

We showed in Sect.~\ref{sec:radioXray} that the wiggles seen in the radio--X-ray correlation of \gx can be explained by simple changes in the disk structure. Within our model, these wiggles are a simple consequence of variations of the disk radiative efficiency with the disk accretion rate. However, in our study, the radio emission along the cycles is computed using three simplifying assumptions. First, we use $b=0.3$ throughout the entire study (\ie, assuming that $30\%$ of the JED power is funneled in the jets). While it would be more physical to assume $b$ to be a function of the physical structure of the accretion flow, this choice is justified both physically \citep{Petrucci10} and spectrally \citep{Marcel18a}. Second, a phenomenological expression for the radio flux $F_R(\dot{m}_{in},r_J)$, see Eq.~(3) and discussion in paper~IV. Third, we used a unique normalization factor $\tilde f_R= 4.5 \times 10^{-10}$ for all cycles, allowing us to qualitatively reproduce the radio--X-ray correlation (see paper~V). There is, however, no physical reason for such a value to be unique. A deeper examination of the radio emission light curves (Figs.~2 and 3 in paper~V) reveals that a better tuning could be done in order to properly describe the radio. In particular, there might be a possible difference between the rising states and the decaying states, or even changes within a given outburst phase. Assuming a function $F_R$ with a unique and constant $\tilde{f}_{R}$ might have generated biases in our results, especially at very low flux when the X-ray spectral shape is not enough to constrain \rj. This is an important question, especially since recent studies found possible differences in both the iron line profiles \citep[][and references therein]{WangJi18, 2020ApJ...899...44W} and the radio--X-ray correlation \citep[][but see Sect.~\ref{sec:radioXray}]{2018MNRAS.481.4513I}. We will address this issue in greater detail in a forthcoming paper \citep{2022A&A...657A..11B}.

\section{Discussion and conclusion}
\label{sec:Ccl}

We use the results from our previous work \citep{Ferreira06, Marcel18a, Marcel18b, Marcel19} and compare the temporal evolution of the two dynamical parameters \rjt and \mdotint . We show that, within the JED-SAD paradigm, their temporal evolution vary significantly between outbursts (see Sect.~\ref{sec:timevol}). However, when illustrated in a \rjdemdot diagram, the path followed shows clear similarities for all considered outbursts of \gx : a fingerprint. While expected, this is an important result and the physical interpretation of this fingerprint will be proposed in a following paper.
Moreover, we can estimate the different processes of energy dissipation in the accretion flow: radiation ($P_{cool}$), advection ($P_{adv}$), and ejection ($P_{jets}$). We thus use the model to estimate each of these powers for all the observations in the fingerprint. We show that three major outcomes can be drawn from this study.

First, the concept of radiative efficiency can be misleading. In particular, we show that within our framework there are two major slopes in the correlation between the X-ray luminosity and the accretion rate: $L_X \propto \dot{m}_{in}$ at low-luminosity and $L_X \propto \dot{m}_{in}^2$ at high luminosity. In our model, we explain these different slopes by the changes observed in \radeff . These two portions have been observed in the past and labeled as radiatively efficient and radiatively inefficient \citep[see, e.g.,][]{Coriat11}. Within our framework we show that, for $\dot{m}_{in} > 10^{-2}$, the disk is in fact never radiatively inefficient because the accretion flow always radiate more than $15-20\%$ of its accretion power. We also argue that the energy band chosen can have a key impact on the slopes obtained between $L_X$ and $\dot{m}_{in}$, often making any estimate unreliable.

Second, we show that there are two different types of hard state spectra, associated with the thermal state of the accretion flow. At low luminosity, the accretion flow is optically thin and geometrically thick, and the total spectrum is that of a simple power law with undetected cutoff: the thick disk regime. At high luminosity, the thermal structure is optically and geometrically slim, and the total spectrum is very similar to the previous power law, although this time with a visible cutoff around $50-100$\,keV: the slim disk regime. In the JED-SAD paradigm this transition unavoidably happens during the rise in the hard state due to the structure of the accretion flow. We thus predict that a similar transition should be observed in other objects or outbursts. We also show that this transition between thick and slim disk regimes is consistent with a change in the evolution of the power-law index $\Gamma_{PL}$, as already observed in \citet{2011MNRAS.417..280S}.

Third, the evolution of the radiative efficiency during the hard state has an important impact on the radio--X-ray correlation. 
While the radio is thought to be linked to the jets structure, in our modelling the X-rays originate from the accretion flow. In consequence, changes in the accretion flow radiative efficiency have a direct impact on the X-rays, but not\footnote{This is true under the assumption that the local disk structure (\eg, vertical extension) has no impact on the production of jets. While this is false in general \citep{Ferreira97}, it was a working assumption in this series of papers and we remain consistent here.} on the radio. In the radio--X-ray correlation, the changes in X-ray \radeff translate into two different slopes separated by a sort of plateau, see Fig.~\ref{fig:Kolj3}. When all hard states are considered, we retrieve the radio--X-ray correlation $L_{\mathrm{Radio}} \propto L_{\mathrm{X}}^{0.6}$ \citep{Corbel13}. When only the most luminous hard states are considered, however, the correlation follows a much steeper slope $L_{\mathrm{Radio}} \propto L_{\mathrm{X}}^{> 1}$, up to $L_{\mathrm{Radio}} \propto L_{\mathrm{X}}^{\approx 1.5}$ depending on the energy ranges considered. Such steep slopes have been associated with a second group of X-ray binaries, labeled outliers, thought of as having a different behavior \citep[see, e.g.,][]{2014MNRAS.440..965H}. We provide here a possible alternate explanation: these outliers could simply be observed at higher accretion rates where the disk radiative efficiency evolves in the slim disk state, rather than the usual thick disk state. Extending this work to other sources, and especially to outliers, should provide valuable constraints and allow this issue to be firmly addressed.

\begin{acknowledgements}
    We would like to thank the referee for their useful and thoughtful comments that helped improve the manuscript. We are thankful to the European Research Council (ERC) for support under the European Union’s Horizon 2020 research and innovation programme (grant 834203; G.M. and C.S.R.), as well as the UK Science and Technology Facilities Council (STFC) for support under Consolidated Grant ST/S000623/1 (C.S.R.). This work was initiated under funding support from the french research national agency (CHAOS project ANR-12-BS05-0009, http://www.chaos-project.fr), the \textit{centre national d'études spatiales} (CNES), and the \textit{programme national des hautes énergies} (PNHE). A.M. acknowledge a financial contribution from the agreement ASI-INAF n.2017-14-H.0, from the INAF mainstream grant (PI: T. Belloni, A. De Rosa) and from the HERMES project financed by the Italian Space Agency (ASI) Agreement n. 2016/13 U.O.. This research has made use of data, software, and/or web tools obtained from the high energy astrophysics science archive research center (HEASARC), a service of the astrophysics science division at NASA/GSFC. All the figures in this paper were produced using the \textsc{matplotlib} package \citep{plt}.
\end{acknowledgements}

\bibliographystyle{aa} 
\bibliography{Research.bib}

\end{document}